\documentclass[%
reprint,
11pt,
nofootinbib,
amsmath,amssymb,
aps,
]{revtex4-2}

\emergencystretch=\maxdimen
\usepackage[utf8]{inputenc}
\usepackage[T1]{fontenc}
\usepackage{fontsize}
\changefontsize[11]{10}
\usepackage{graphicx}%
\usepackage{dcolumn}%
\usepackage{bm}%
\usepackage{ragged2e}
\usepackage{amsmath,amssymb} %
\usepackage{mathrsfs}
\usepackage{siunitx}
\usepackage{amsfonts}
\usepackage{relsize}
\usepackage{pgfplots}
\usepackage{floatrow}
\usepackage{float}
\usepackage{caption}
\usepackage{subfig}
\usepackage{xcolor}
\usepackage[symbol]{footmisc} %
\definecolor{Green}{HTML}{167425}
\definecolor{azure}{HTML}{003FFF}

\usepackage[colorlinks=true,linkcolor=Green,citecolor=blue]{hyperref}%

\def\fig{Figure~}

\def\eqns{Eqs.~}

\usepackage[normalem]{ulem}

\def\IB#1{\boldsymbol{#1}} %

\usepackage{lineno}
\begin{document}

\title{
    Colloid recovery from porous structures under ambient flow: enhanced extraction via phoretic and osmotic mechanisms}
\author{Jitendra Dhakar}%
\author{Kapil Upadhyaya}
\author{Akash Choudhary}
\email{\textcolor{black}{Corresponding author:} \textcolor{black}{achoudhary@iitk.ac.in}}
\address{Department of Chemical Engineering, Indian Institute of Technology Kanpur, 208016, India}

    \begin{abstract} 
\noindent
\textbf{Abstract.} Chemical gradients are widely employed to enhance particle transport in porous media, such as laundry detergency and enhanced oil recovery. 
Diffusiophoresis and diffusioosmosis refer to the movement of colloid and movement of near-surface fluid in response to electrolyte gradients, respectively. These mechanisms play a crucial role in colloid and drug transport in constricted regions where bulk transport is infeasible. 
Our earlier work [Tiwari et al., Langmuir 41, 18583 (2025)] has shown that phoretic and osmotic transport in dead-end micro-pores can be controlled by orienting salt gradients into or out of the pores; however, the extent to which this orientation influences large-scale spatiotemporal patterns and colloid extraction is not thoroughly explored. 
In this work, we study the phoretic and osmotic colloidal extraction from porous structure exposed to an ambient flow.
We study the influence of solute gradient orientation, namely solute-out (i.e., initially salt-rich yarn) and solute-in (i.e., initially salt-free yarn) modes.
The two-dimensional porous structure is made of a number of pillars/fibers arranged in a lattice ordered hexagonal packing with equal spacing. 
The results from finite-element simulations show that phoretic colloidal extraction exhibits a qualitatively distinct behavior in the two modes. 
Specifically, in the solute-out mode, colloids are preferentially extracted from the peripheral region of the porous structure, whereas in the solute-in mode, extraction predominantly occurs from the stagnant core region. 
Diffusioosmotic slip on the internal surface of pillars/fibres further amplifies extraction in both modes, with a relatively larger enhancement in the solute-in mode due to internal spatiotemporal flow patterns.
Beyond demonstrating the sensitivity of osmotic transport in porous media, these insights can guide enhanced membrane filtration, laundry detergency, and enhanced oil recovery.
    \end{abstract}
    \maketitle

\section{Introduction}
\begin{figure}[t]
    \centering
\includegraphics[width=8cm]{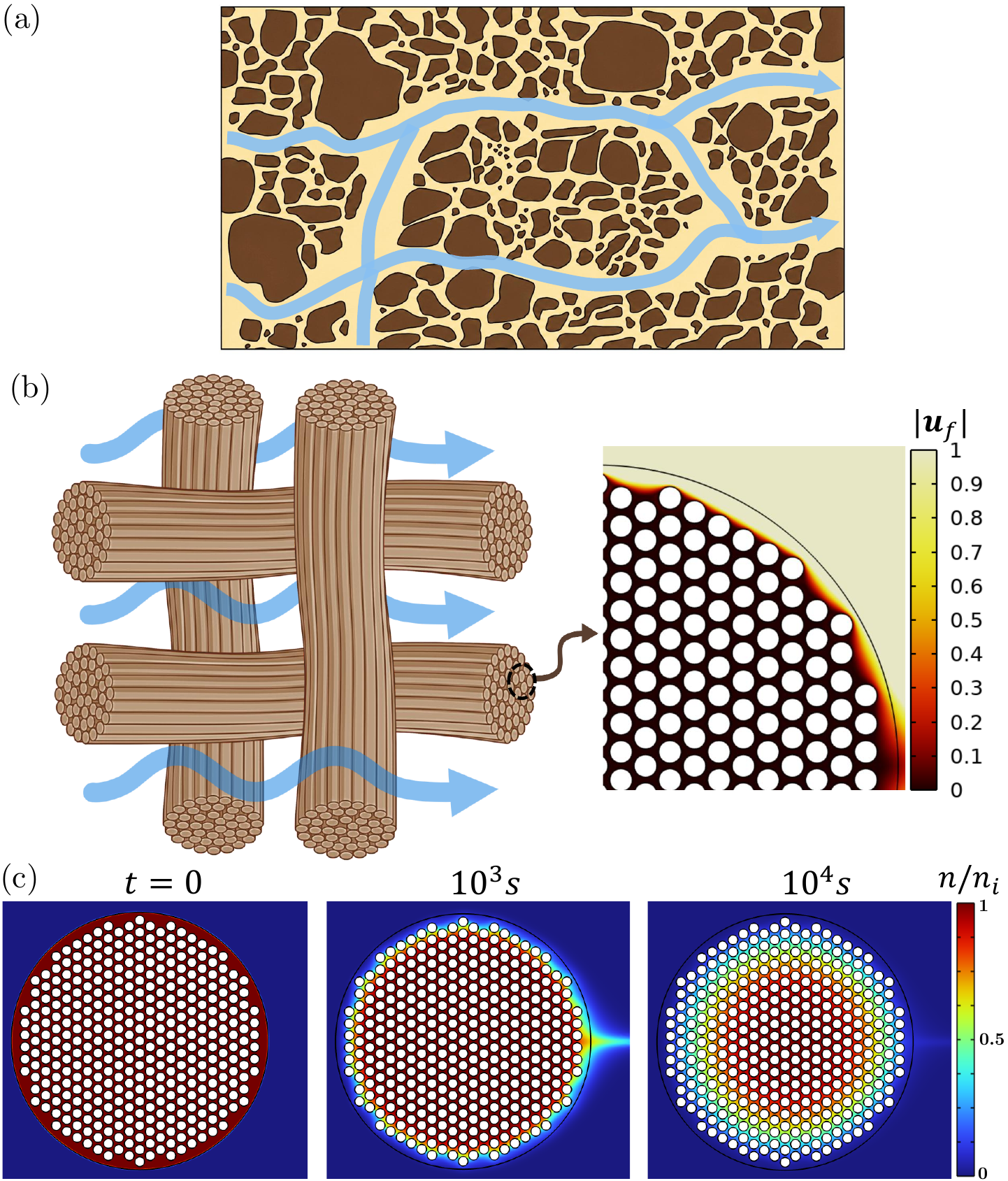}
    \caption{Schematic representation of (a) oil trapped in the stagnant low-permeability regions of fractured porous rocks and (b) dirt colloids trapped in the textile porous yarn. The external flow around such structures is unable to trigger advection, and hence colloids are extracted via slow colloidal diffusion. $|\IB{u}_f|  =  \sqrt{\IB{u}_{fx}^2 + \IB{u}_{fy}^2}$ denotes the velocity intensity.
    (c) Numerical results showing slow extraction of colloids (approximately 0.5 $\mu$m diameter) that are trapped inside a cylindrical porous structure ($R=4 \times 10^{-4}$m) subjected to external flow of Re$=$0.02. See Table \ref{tab:para} for further details of parameters.
    }
    \label{Fig:beta1}
\end{figure}

\noindent 
The transport of nano and microscale particles and emulsions through porous media alongside nutrients or solutes is essential for applications in oil recovery \citep{lake1988enhanced,lager2008losal}, groundwater remediation \citep{corapcioglu1993colloid} and water filtration \citep{shin2017membraneless,miele2019stochastic}.
Typically, such suspensions are transported by applying pressure gradients. At micrometer length scales, however, this approach becomes prohibitively expensive as the gradients required to drive flow through confined pores grow impracticably large \citep{bird2002phenomena}.
Porous structures in both natural and engineered settings, frequently feature dead-end pores or stagnant zones where fluid advection is negligible, leaving transport to be dominated by diffusion \citep{warmoeskerken2002laundry,wu2016multiphase}.
Broadly, these challenges associated with stagnant zones arise wherever flow must navigate around solid obstacles within the pore network. 
During chemical enhanced oil recovery in heterogeneously permeable media, emulsified oil can get trapped within low-permeability pockets, whereas the external sweeping flow is short-circuited through high-permeability regions or fractures \citep{lake1988enhanced,wu2016multiphase}; \fig~\ref{Fig:beta1}(a) illustrates this.
In the context of textile laundry, this issue occurs more intensely, where dirt particles are trapped in the inter (100 $\mu$m) and intra-yarn (10 $\mu$m) spaces (see Fig. \ref{Fig:beta1}b). 
During a typical washing operation, external advection generates pressure drop of $\sim$ $10^3-10^4 \, N/\text{m}^2$ across few centimetres of bundled fabric, whose permeabilities of macro- and micro-pores are roughly $10^{-11}$ and $10^{-14}$ $\text{m}^2$, respectively \cite{van1987hydrodynamics,warmoeskerken2002laundry}. Assuming isotropic and homogeneous structure of macro- \& micro-pores, Darcy's law can be used to estimate the inter and intra-yarn Reynolds number ($U_\infty R/\nu$) as $\sim 0.1$ and $10^{-5}$, respectively. 
Consequently, the intra-yarn local P\'eclet number ($U R / D_p$) remains well below unity despite the low Brownian diffusivity of colloidal dirt.
Thus, the transport of dirt in the intra-yarn spaces is governed by slow Brownian diffusion, as Figure 1 (c) shows the 2D simulations of flow around an $800 \mu$m diameter cylindrical porous structure with micron sized colloids trapped within it.
Despite their heterogeneity, these systems share a common transport bottleneck: interconnected, tortuous stagnant zones largely bypassed by an external sweeping flow; this forms the focus of current study.

Recent studies of flow through porous media have shown that using salt or surfactant gradients can trigger a `diffusiophoretic' motion that can enhance particle \citep{shin2016size,shin2018cleaning,jotkar2024impact,somasundar2023diffusiophoretic} and drop mobilization \citep{yang2018diffusiophoresis,park2021microfluidic,duong2026salt} in dead-end pores and porous media. 
Succinctly, diffusiophoresis (DP) is a physicochemical transport mechanism by which colloidal particles migrate in response to electrolyte concentration gradients.
When an electrolyte dissociates, its constituent ions diffuse at different rates, giving rise to a local electric field that accompanies the ionic flux; the colloid interacts with both and migrates with a drift velocity proportional to $\nabla \log c$ \citep{prieve1984motion,anderson1989colloid}. 
The net magnitude and direction depends on the solute-surface interactions and are captured by mobility parameter. 
Under similar circumstances, pore wall surfaces have tendency to exhibit a slip-driven diffusio-osmotic (DO) flow that aids colloidal mobility
\citep{derjaguin1972capillary,ault2025physicochemical}.

\citet{shin2018cleaning} demonstrated that textile materials can be efficiently cleaned by exploiting sharp surfactant gradients. 
Through pore-level experiments and simulations, they showed that such gradients drive diffusiophoretic extraction at rates two orders of magnitude faster than in non-gradient systems.
However, most prior studies on DP and DO have been largely confined to single-pore systems \citep{shin2016size,ault2017diffusiophoresis,alessio2022diffusioosmosis}. 
Very recently,  \citet{jotkar2024impact,jotkar2024diffusiophoresis} characterized the enhancement offered by DP in flows through porous media.
They reported that DP alters particle dispersion and imparts non-Fickian signature in colloidal dispersion and residence times.
\citet{chu2019advective} analytically examined diffusiophoretic spreading of colloids driven by a diffusing solute patch in an unbounded domain, and showed that the ratio of the diffusiophoretic mobility to the solute diffusivity determines whether colloids undergo super-diffusion, transiently arrested spreading, or regular diffusion.

Diffusiophoretic exchange between stagnant regions and an adjacent flow has received further attention. \citet{battat2019particle} showed that colloids flowing past dead-end pores can undergo pore-to-pore hops before capture. 
\citet{akdeniz2023diffusiophoresis} showed that diffusio-osmosis with a concentration-dependent zeta potential modulates particle transport into dead-end channels, while \citet{migacz2024enhanced} showed that time-dependent and oscillatory inlet solute concentrations can tune injection and withdrawal efficiency. 
Recently, we showed that colloid mobilization in dead-end pores is highly sensitive to direction of imposed gradients \citep{tiwari2025influence}. Solute-out mode (gradient oriented towards dead end) or solute-in mode (gradient oriented towards pore mouth): colloidal withdrawal in solute-out mode is faster and shallower, whereas the solute-in mode enables deeper withdrawal.

Through microfluidic experiments and simulations of flow through ordered and disordered porous media, \citet{alipour2026diffusiophoretic} showed that DP can enhance (or impair) colloidal extraction via cross-stream migration from dead-end zones towards preferential flow pathways (or away from them).
More recently, \citet{li2026solute} found that sharp solute fronts extract colloids faster at early times, whereas diffuse fronts achieve higher removal at late times, and \citet{pujari2026diffusiophoretic} showed that phoretic exchange between slow and fast streamlines governs colloid dispersion in porous media. 

While diffusiophoresis in flow \textit{through} porous media has been analyzed in these aforementioned studies, flow \textit{over} a finite porous inclusion remains largely unexplored.
The interplay between external flow and the coupled dynamics of salt and colloid gradients raises questions that this work seeks to address:
(i.) How efficient is diffusiophoresis for extracting colloids from the porous media under external ambient flow?
(ii.) How do the two orientational modes of solute gradient affect the spatiotemporal patterns and efficiency?
(iii.) What is the impact of diffusio-osmotic slip at the confining walls of pores? 

In what follows, we present the continuum framework employed to predict and obtain qualitative insights into colloidal withdrawal from porous media under ambient flow. 
The geometry considered here represents an idealized transverse cross-section of a textile yarn subjected to a uniform external flow perpendicular to its length.
Specifically, in Section \ref{sec:ii}, we present the mathematical model for solute and colloid transport in the porous structure with ambient flow. 
In Section \ref{sec:iii}, we discuss the influence of internal osmotic flow and its distinct spatial signature in the two modes. Lastly, in Section \ref{sec:iv}, we highlight the key findings of this work.

\begin{figure}[h]
    \centering
    \includegraphics[width=7cm]{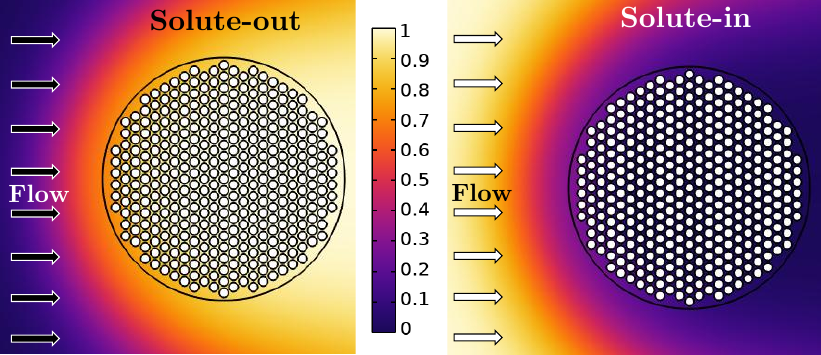}
    \caption{Schematic representation of two solute gradient modes: S-O ($\beta=0.01$) and S-I ($\beta=100$), where $\beta=c_{x=0}/c_{t=0}$ represents the strength of the inward or outward solute gradient.}
    \label{Fig:geometry}
\end{figure}

\section{Mathematical Model}
\label{sec:ii}
\noindent We study the extraction of colloids from a porous two-dimensional structure subjected to ambient flow that flows between a sufficiently spaced array of such structures.
The ambient flow around is set with the average inflow velocity $U_\infty$. The obstruction to flow generates disturbance flow fields that resembles an unbounded two-dimensional flow past a cylinder \citep{proudman1957expansions,batchelor2000introduction} (see \S 1 in Supplementary material). 
\fig\ref{Fig:geometry} shows that the surrounding flow can set up electrolytic gradients in two modes. (i.) Solute-out mode: the porous structure pre-saturated with electrolytic solution is subjected to ambient flow of electrolyte-free water; (ii.) Solute-in mode: salt-free porous structure is subjected to ambient flow of electrolytic solution.
The structure of radius $R$ resembles a yarn that is made up of a hexagonal packing of bundles of fibers (radius, $R_f$) with surface-to-surface gap between fibers $S_f=7.5$ $\mu$m. The yarn contains a number of fibers $N_f=367$, and the porosity $\phi=1 - (A_{fibers}/A_{yarn})=1-(N_f \pi R_f^2)/(\pi R^2)$.
Tortuosity is estimated from the Bruggeman relation $\tau_u = \phi^{-1/2}$ \cite{tjaden2016origin}; however, since the pore geometry is fully resolved, neither $\phi$ nor $\tau_u$ enters the simulations as an input parameter.
In our simulations, this structure is placed in an outer domain, with length ($L$) and width ($W$) set to $20R$ each.

The model is governed by the Navier-Stokes equations (\eqns\ref{eq:1a} and \ref{eq:1b}) for fluid flow, and the convection-diffusion equations (\eqns\ref{eq:1c} and \ref{eq:1d})  for solute and colloid transport, respectively:
\begin{gather}
 \rho ( \frac{\partial \mathbf{u}_f}{\partial t} + \mathbf{u}_f \cdot \nabla \mathbf{u}_f ) = -\nabla p + \mu \nabla^2 \mathbf{u}_f , \label{eq:1a} \\
\nabla \cdot \mathbf{u}_f = 0, \label{eq:1b} \\
  \frac{\partial c}{\partial t}+ \nabla \cdot [\mathbf{u}_f c] =  \mathcal{D}_\text{s} \nabla^2 c ,  \label{eq:1c} \\
\frac{\partial n}{\partial t} +\nabla \cdot [(\mathbf{u}_f + \mathbf{u}_p) n] =  \mathcal{D}_\text{p} \nabla^2 n ,\label{eq:1d}
\end{gather}
where $\rho$, $\mu$, $p$, $\mathbf{u}_f$, $c$, $\mathcal{D}_\text{s}$, $n$, $\mathcal{D}_\text{p}$, $t$, and $\mathbf{u}_p$ denote the medium density, viscosity, pressure, flow velocity, solute concentration and diffusivity, colloid concentration and diffusivity, time, and diffusiophoretic velocity $(\Gamma_p \nabla \ln c)$ of colloid, respectively. 
Assuming that the electrolyte is symmetric and the concentration gradient is weak on the scale of the colloid size (\textit{i.e.,} the gradient length scale greatly exceeds the particle radius), the phoretic mobility ($\Gamma_p$) takes the following form \citep{prieve1984motion}:
\begin{gather}
	\Gamma_p = \frac{\varepsilon}{2\mu}\left(\frac{k_\text{B}T}{Ze}\right)^2 \left[2\mathcal{B}_s\frac{Ze\zeta_p}{k_BT} + 8 \ln cosh\left(\frac{Ze\zeta_p}{4k_\text{B}T}\right)\right],
	\label{eq:GammaP}
\end{gather}
where $\varepsilon$, $k_B$, $T$, $Z$, $e$, $\mathcal{B}_s\left(=\frac{\mathcal{D}_{+}-\mathcal{D}_{-}}{\mathcal{D}_{+}+\mathcal{D}_{-}}\right) $, and $\zeta_p$ represent the medium permittivity, Boltzmann constant, absolute temperature, ionic valency, elementary charge, and diffusivity contrast between cation and anion, colloid zeta potential, and particle zeta potential respectively.

As the solute front passes the yarn, the conditions outside the porous structure are $c=c_i$ and $n=0$ for the solute and colloid concentrations, respectively.
Inside the domain, we have $c=c_i$ and $n=n_i$ uniformly at $t=0$.
The inlet boundary conditions of the outer domain are given by $c=\beta c_i$, where $\beta =0.01$ (S-O) and $\beta =100$ (S-I) mode; the inlet colloid concentration is $n=0$. 
The diffusive fluxes of solute and colloid are assumed to be zero on the walls of the structure, outer domain walls and outlet boundary. 
The outlet boundary of the simulation domain is open to the ambient, i.e., $p=0$. 
Furthermore, we account for the diffusio-osmotic slip ($ -\Gamma_w \nabla \log c $) on the inner walls of the porous cylinder due to finite wall mobility (stemming from finite surface charge). We will show that the DO slip generates an additional internal flow ($\IB{u}_f$) and contributes significantly to particle withdrawal. 
Detailed descriptions of the boundary conditions are provided in \S 3 of the Supplementary material.
The mathematical model consisting of governing equations (i.e., \eqns\ref{eq:1a} to \ref{eq:1d}) with relevant boundary conditions, is solved using the finite element method (FEM) based {\small{COMSOL}} Multiphysics (V.6.3) to obtain the solute and colloid concentration, and velocity profiles in the porous structure.
The methodology involves fully coupled {\small{PARDISO}} linear and Newton's non-linear time-dependent solvers. 
An unstructured, free triangular mesh with boundary layers on confined pore/fiber walls is used to discretize the computational domain. A mesh independence study (see Appendix A and \S3 in Supplementary material) is performed by evaluating the total number of colloids (per unit length) for successive refinements of the mesh.

\begin{table}[b]
			\centering
			\caption{Domain dimensions and parameters used in the numerical simulations.}\label{tab:para}
			\vspace{2mm}
			\scalebox{1}
			{
				\begin{tabular}{lll}
					\hline
                    $\phi$ & Porosity & 0.48 \\ 
                    $\tau_u$ & Tortuosity & 1.43 \\
                    $\mathcal{D}_s$ & Solute ambipolar diffusivity & $10^{-9}$ m\textsuperscript{2}/s  \\
                    $\beta$ & Solute concentration gradient & 0.01, 100   \\
                    $\mathcal{D}_p$ & Colloid diffusivity & $10^{-12}$ m\textsuperscript{2}/s  \\
                    $\Gamma_p$ & Colloid mobility & $\pm 2\times 10^{-10}$ m\textsuperscript{2}/s  \\
                    $\Gamma_w$ & Fiber walls mobility & $\pm 5\times 10^{-10}$ m\textsuperscript{2}/s  \\
                    $U_\infty$ & Ambient flow velocity &  $5\times 10^{-5}$ m/s  \\
                    \hline
				\end{tabular}
			}
		\end{table}
As a physically motivated reference case, these are chosen to reflect the typical microstructure of textile yarn in laundry processes \citep{warmoeskerken2002laundry}: yarn radius $R = 400\,\mu$m, fibre radius $R_f = 15\,\mu$m, and fibre spacing $S_f = 7.5\,\mu$m.
Colloids of size $0.5\,\mu\text{m}$ have a diffusivity of $\mathcal{D}_p = 10^{-12}\,\text{m}^2/\text{s}$, estimated from the Stokes-Einstein relation. 
We have considered wall mobility ($\Gamma_w = 5 \times 10^{-10}$ m\textsuperscript{2}/s) and particle mobility ($\Gamma_p = 2 \times 10^{-10}$ m\textsuperscript{2}/s) that are modeled after a system comprising polyethylene terephthalate (PET) fibers and weakly charged colloidal particles immersed in aqueous NaCl electrolyte solution. Under these conditions, the diffusivity difference factor for NaCl is $\mathcal{B}_s\approx -0.21$. Experimental literature regarding the electrokinetic properties of textiles demonstrates that PET fibers exhibit highly negative zeta potentials, typically ranging from $-50$ mV to $-60$ mV in dilute monovalent salt solutions \citep{grancaric2005electrokinetic,jacobasch1985problems}. \citet{han2022fibrous} have depicted ($\zeta_w \approx -58 \text{ mV}$)  for PET fibres in 10 mM NaCl.
Using the standard formulation for diffusioosmotic mobility, substituting a wall zeta potential of $\zeta_w = -58 \text{ mV}$ yields a theoretical wall mobility of $\Gamma_w \approx 5.0 \times 10^{-10}$ m\textsuperscript{2}/s. Similarly, for a moderately charged particle ($\zeta_p \approx -30 \text{ mV}$) \citep{akdeniz2023diffusiophoresis,kar2015enhanced} in the same electrolyte, the resulting particle mobility is $\Gamma_p \approx 2.0 \times 10^{-10}$ m\textsuperscript{2}/s.
Hence, our chosen ratio of $|\Gamma_w / \Gamma_p| = 2.5$  reflects a plausible scenario for the strongly charged structural walls and moderately mobile particles modeled in our geometry.
In contrast, unilamellar lipid vesicles ($\zeta_p = -90\,\text{mV}$) suspended in SDS ($\mathcal{B}_s = 0.55$) exhibit a negative mobility $\Gamma_p = -0.2 \times 10^{-9}\,\text{m}^2/\text{s}$ \citep{shin2017low}. 
By analogy, we also consider both positive and negative phoretic mobilities for the fibres (walls) constituting the porous structure \citep[p. 337]{luxbacher2020electrokinetic}.
However, we note that systems with opposite zeta potentials of particles and walls are challenging to study experimentally due to electrostatic fouling \citep{breite2016critical}.
The dimensional simulation parameters are listed in Table~\ref{tab:para}.

\section{Results and Discussion}
\label{sec:iii}

\subsection{Characterizing and validating transport in no-gradient control case}
Before quantifying the extraction efficiency of phoretic and osmotic phenomena, we analytically examine the dynamics of no-gradient colloidal transport \textit{i.e.,} the control case ($\beta=1$) where both phoretic and osmotic effects are absent.
The total number of colloids (per unit length) at time $t$  are evaluated as: $N(t) =\int_0^R 2\pi r n(r,t)\,dr$, and at $t=0$: $N_i \sim n_i \pi R^2$.
Figure \ref{Fig:control} shows how the normalized average concentration decreases over time as colloids are withdrawn by external convective flux. To rationalize this decay, we derive analytical expressions in two temporal limits; assuming the ambient to be a perfect sink with infinitely high mass transfer coefficient, we assume a homogeneous Dirichlet boundary condition
$n(R,t)=0$ for the subsequent derivation.

\noindent
\textbf{Short-time scaling} ($ t \ll R^2/\mathcal{D}_p $).
At early times, diffusion only removes a thin layer of colloids from the outer regions of porous structure:
$\delta(t) \sim \sqrt{\mathcal{D}_pt},$
with $\delta \ll R$. Since the system can be modeled locally planar, we write the well-known semi-infinite solution as $n(x,t) = n_i \,\mathrm{erf}\!\left({x}/{\sqrt{4\mathcal{D}_pt}}\right)$, where $x = R-r$.
Integrating in $x$ from $0$ to $\infty$ yields net withdrawn colloids as
\begin{equation}
    N_i - N(t)  = 2 \pi R \mathlarger{\int}_{0}^{\infty}   n_i {\small \left[ 1- \text{erf}\left( \frac{x}{\sqrt{4\mathcal{D}_p t}} \right) \right] }   dx.
\end{equation}
Rearranging, we get:
\begin{equation}\label{eq:STS}
    \frac{N(t)}{N_i} = 1-\frac{4}{\sqrt{\pi}} \sqrt{\frac{t}{t_p}},
\end{equation}
where $t_p = R^2/\mathcal{D}_p$ is the diffusive time scale of colloids.

\noindent
\textbf{Long-time scaling} ($ t \gg R^2/\mathcal{D}_p $): 
Here, we seek separable solutions of the form $n(r, t) = \phi(r) e^{-\lambda t}$, whose substitution in the diffusion equations yields the zero-order Bessel equation:
\begin{equation}
    \frac{d^2\phi}{dr^2} + \frac{1}{r}\frac{d\phi}{dr} + \frac{\lambda}{\mathcal{D}_p}\phi = 0, \quad \phi(R) = 0.
\end{equation}
This yields axisymmetric solution that is superposition of the following modes: $\phi_n(r) = A_n J_0(r\sqrt{\lambda_n/\mathcal{D}_p})$, 
where $ \lambda_n =  {\mathcal{D}_p} \, x_{0,n}^2/{R^2},$ and $x_{0,n}$ are the roots of the $J_0$ Bessel function. 
At long times, we can expect the first-term ($\phi_1$ with $x_{0,1} \approx 2.405$) to dominate the solution:
\begin{equation}
    n(r, t) \sim A_1 J_0\left( \frac{x_{0,1} r}{R} \right) \exp\left( -x_{0,1}^2 \frac{\mathcal{D}_pt}{R^2} \right).
\end{equation}
Therefore the total number of colloids per unit length scales as
\begin{equation}\label{eq:LTS}
    \frac{N(t)}{N_i} = 0.692 \exp\left( -x_{0}^2 \,{t}/{t_p} \right).
\end{equation}

In the extremely low Reynolds number limit, Fig. \ref{Fig:control} shows that Eqs.~(\ref{eq:STS}) and (\ref{eq:LTS}) closely predict the withdrawal dynamics obtained from simulations. 
The slight discrepancy observed at late times arises from the tortuosity of the porous domain: unlike the idealized radial transport assumed in the analytical model, colloids in the simulation traverse effectively increased path length.
At higher ambient Reynolds numbers ($Re$), the convective transport near the pore exterior leads to a faster departure from the early-time predictions, which assume a purely diffusive transport within the pore.
\\

\begin{widetext}

\begin{figure}[t]
    \centering
    \includegraphics[width=0.4 \textwidth]{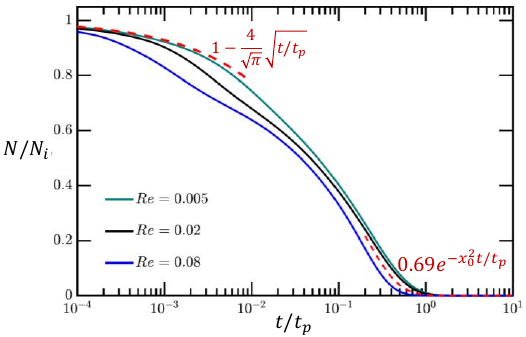}
    \caption{\small \textbf{Colloid extraction for no-gradient case.} Comparison of total number of colloids (per unit length) variation with time for numerical (solid lines) and analytical results (dashed line) for early time and late time scaling, respectively. Here $t_p=R^2/\mathcal{D}_p$ ($=1.6\times 10^5$) denotes colloid diffusion time and $Re=\rho U_\infty R/\mu$. 
    }
    \label{Fig:control}
\end{figure}

\begin{figure*}[b]
    \centering
    \includegraphics[width=\textwidth]{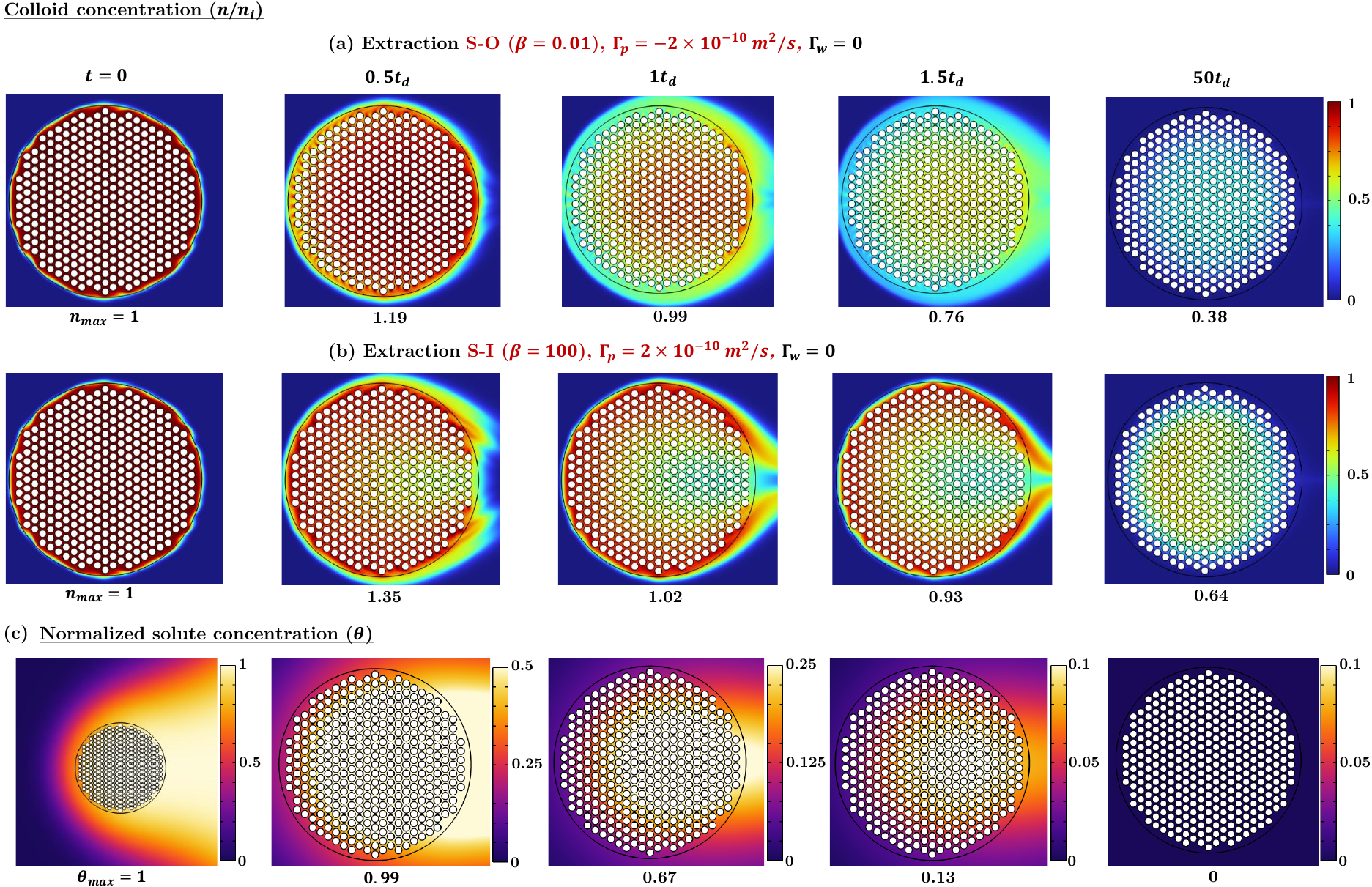}
    \caption{\small \textbf{Colloid and solute profiles inside the porous media.} Colloid concentration variation with time (in units of $t_d$) for (a) solute-out mode ($\beta=0.01$), and (b) solute-in mode ($\beta=100$), respectively. (c) Variation of normalized solute concentration ($\theta=(c-c_\infty)/(c_i-c_\infty)$) with time. Parameters: $\mathcal{D}_s = 10^{-9}\,\mathrm{m}^2/\mathrm{s}$, $\mathcal{D}_p = 10^{-12}\,\mathrm{m}^2/\mathrm{s}$, $\Gamma_p = \pm 2\times 10^{-10}\,\mathrm{m}^2/\mathrm{s}$, $R=400 \mu m$ (yarn radius), $R_f=15 \mu m$ (fiber radius), $S_f=7.5 \mu m$ (spacing), $\phi=0.48$ (porosity), $\tau_u=1.43$ (tortuosity). Here, time $t=0$ is considered the time ($t_i=L_i/U_\infty=32s$) at which the solute front reaches the porous yarn in the outer domain, where $L_i$($=0.25L - R$) is the distance between outer domain inlet and the porous yarn.
    }
    \label{Fig:DP_profiles}
\end{figure*}

\end{widetext}

\clearpage
\clearpage

\subsection{Phoretic extraction of colloids from porous media}  
In this section, we build on the results for the $\beta = 1$ regime in \S III.A by incorporating the diffusiophoretic response in the solute-out ($\beta < 1$) and solute-in ($\beta > 1$) modes (\fig\ref{Fig:DP_profiles}).
Here, we have considered both positive and negative colloid mobility, which can result either from a positive or negative zeta potential, or from the difference in ionic diffusivity ($\mathcal{B}_s=\frac{\mathcal{D}_{+}-\mathcal{D}_{-}}{\mathcal{D}_{+}+\mathcal{D}_{-}}$) \citep{prieve1984motion}. 
Since this study focuses on phoretic colloid extraction, we consider negative colloid mobility for the S–O mode and positive mobility for the S–I mode (i.e., extraction-compatible pairing). 
Physically, negative mobility drives the colloid away from regions of higher solute concentration, whereas positive mobility drives it toward them. 
This sign convention, therefore, distinguishes the outward and inward phoretic motion considered in this study. 
The other pairing triggers inward compaction and is shown in \S 4 in Supplementary material.

Panels (a) and (b) in \fig\ref{Fig:DP_profiles} show the temporal evolution of colloid concentration profiles in units of characteristic solute diffusion time $t_d = R^2/\mathcal{D}_s$, which for the current setup is 160 s (Table \ref{tab:para}).
The first notable feature is that both modes offer significant withdrawal within single $t_d$, whereas bare colloid diffusion offered orders-of-magnitude weaker extraction ($t \gtrsim 100 \,t_d $), as shown in Fig.~\ref{Fig:beta1}(c).
However, the two modes exhibit contrasting extraction dynamics: the S-O mode rapidly ejects colloids from the outer regions of the cylinder before drawing them from the core. 
In contrast, the S-I mode initiates extraction from the inside out, clearing the central core first before gradually releasing the remaining particles from the porous structure. 
Notably, the colloid depleted core is shifted downstream, which correlates with the evolution of the normalized electrolyte concentration $\theta={(c-c_\infty)}/{(c_i-c_\infty)}$ in \fig\ref{Fig:DP_profiles} (c).
The bulk flow advects the solute downstream, resulting in reduced concentration gradients and, therefore, a weaker radial withdrawal in the downstream region (for both modes).

\begin{figure}[t]
    \centering
\includegraphics[width=1 \textwidth]{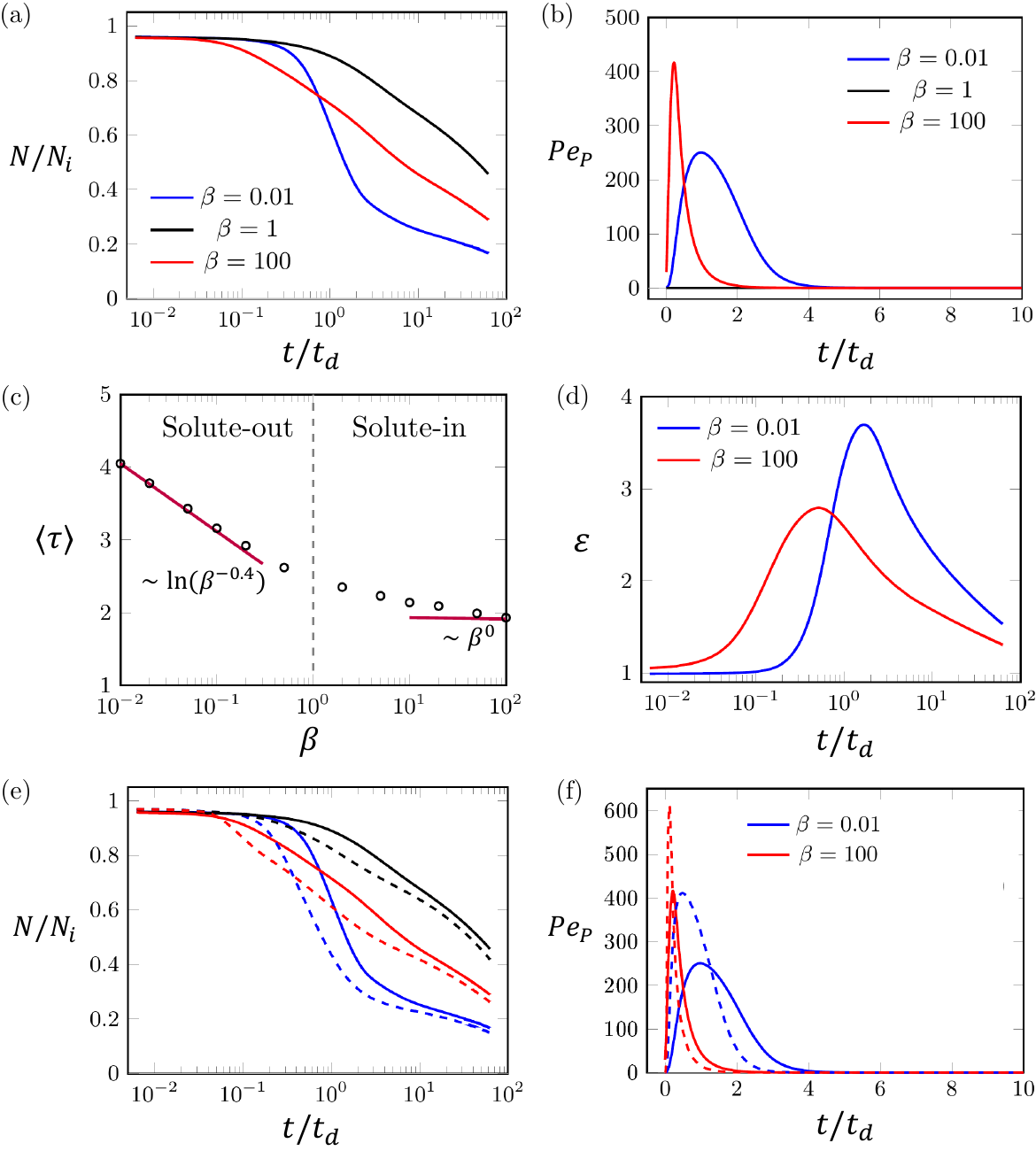}
    \caption{\small \textbf{Quantification of colloid extraction.} (a) Total number of colloids (per unit length) ($N/N_i$) and (b) phoretic Péclet number ($\text{Pe}_P = \langle \mathbf{u}_p \rangle R/\mathcal{D}_p$) variation with time ($t/t_d$). (c) Persistence time variation with solute concentration gradient. (d) Effectiveness ($\varepsilon$) of colloid extraction variation with time. The blue, black and red lines denote S-O ($\beta =0.01$), no-gradient ($\beta =1$), and S-I ($\beta =100$) modes, respectively. 
    (e) Temporal profile of colloid withdrawal for various $Pe_A$. Here $Pe_A = 2 \times 10^4$ (solid lines) and $Pe_A = 8 \times 10^4$ (dashed lines). (f) Temporal variation of $\text{Pe}_P$ for different $Pe_A$. 
    Parameters: $\mathcal{D}_s = 10^{-9}\,\mathrm{m}^2/\mathrm{s}$, $\mathcal{D}_p = 10^{-12}\,\mathrm{m}^2/\mathrm{s}$, $\Gamma_p = \pm 2\times 10^{-10}\,\mathrm{m}^2/\mathrm{s}$.
 }
    \label{Fig:DP_performance_measure}
\end{figure}

\fig\ref{Fig:DP_performance_measure} (a) quantifies how total number of colloids (per unit length) rapidly decays with time for both modes.
Notably, the S-I mode shows accelerated colloid mobilization at early times, whereas S-O mode intensifies extraction at later times.
To understand this contrast, we derive the expression for phoretic Péclet number: $\text{Pe}_P = \frac{\langle {\mathbf{u}_p} \rangle R}{\mathcal{D}_p},$ 
where $\langle \IB{u}_p \rangle$ is the domain-averaged scalar magnitude of the phoretic velocity. This velocity can be written in normalized form as:
\begin{equation}
     \IB{u}_p = \Gamma_p\frac{ \nabla\theta }{\theta  +\frac{c_\infty}{c_i-c_\infty}},
\end{equation}
which, in the limits of S-I and S-O modes, yield the Péclet number as:
\begin{gather}
 \text{Pe}_P = 
 \begin{cases}
     \frac{\Gamma_pR}{\mathcal{D}_p} \langle {\frac{\nabla\theta }{\theta }} \rangle \quad \text{for} \quad c_i \gg c_\infty \quad \text{(S-O)}\\
     \frac{\Gamma_pR}{\mathcal{D}_p} \langle  {\frac{\nabla\theta}{\theta -1}} \rangle \quad \text{for} \quad c_i \ll c_\infty \quad \text{(S-I)}
 \end{cases}\label{eq:Pec}
\end{gather}
Consistent with \fig\ref{Fig:DP_performance_measure}(a), the figure (b) depicts an earlier peak in $\text{Pe}_P$ for S-I mode.
This difference stems from their respective denominators in Eq.(\ref{eq:Pec}) because the dimensionless gradients ($\nabla \theta$) in the numerator are identical for both modes. 
At early times ($t\ll t_d$) \fig\ref{Fig:DP_profiles}(c) shows that $|\theta| \sim 1$; the domain undergoing S-I mode is predominantly devoid of solute, corresponding to $|\theta - 1| \ll 1$. 
Conversely, the S-O mode features a electrolyte-filled domain at early times, yielding a larger denominator.
On the other hand, at larger times ($t \gtrsim t_d$), the denominator of S-O mode is smaller as most electrolyte has diffused out of the domain ($|\theta| \ll 1$). Therefore, we observe S-O dominance at larger times.
We also note that the maximum value of $\text{Pe}_P$ is larger for S-I mode because it dominates at early times, when concentration gradients $(\nabla\theta)$ are largest.
Due to these reasons, \fig\ref{Fig:DP_performance_measure}(a) shows a crossover from S-I mode dominance at early times to S-O mode dominance at later times.

\fig\ref{Fig:DP_performance_measure}(c) shows the overall lifespan or persistence time $\langle \tau \rangle$ for various applied solute gradients $\beta$ (S-O corresponds to $\beta<1$ and S-I to $\beta>1$).
$\langle \tau \rangle$ is defined as the time required for the domain-averaged diffusiophoretic velocity ($\langle \mathbf{u}_p \rangle $) to decrease to $1\%$ of its maximum value.  
For the S-O case, $\langle \tau \rangle$ increases as $\beta$ decreases, consistent with the scaling $\sim \ln(\beta^{-0.4})$ estimated in our earlier study on dead-end pores \citep{tiwari2025influence}.
Whereas in the S-I regime, persistence time decreases for increasing $\beta$ and becomes indifferent to $\beta$. This saturation arises because increasing $\beta$ enhances both gradient and the local concentration at the solute front in a compensating manner, leaving the DP velocity (logarithmic gradient of solute concentration), and hence $\langle \tau \rangle$, effectively unchanged.

 The performance of colloid extraction from porous structure is measured by effectiveness, which is defined as the ratio of total colloid extracted in the presence of a solute gradient and without a solute gradient:
\begin{gather}
\varepsilon (t) = \frac{1-{\langle n/n_i \rangle}_{\beta\neq1}}{{1-\langle n/n_i \rangle}_{\beta=1}}. \label{eq:eff}
\end{gather}
Fig.~\ref{Fig:DP_performance_measure}(d) shows effectiveness measurements consistent with Fig.~\ref{Fig:DP_performance_measure}(a,b): S-I enables rapid colloid extraction, whereas S-O mode persists and achieves better overall extraction in a phoretic lifespan.
Further, the S-O mode exhibits a higher cumulative effectiveness per unit lifetime ($\sim 4t_d$) than the S-I mode, with corresponding values of $3.05$ and $2.25$, respectively (we obtain these estimates by integrating the curve in Fig.~\ref{Fig:DP_performance_measure}(d)).

The influence of ambient flow strength on colloid extraction is depicted in the \fig\ref{Fig:DP_performance_measure}(e). The ambient flow is characterized by the ambient P\'eclet number $Pe_A = U_\infty R/\mathcal{D}_p$, which compares colloid advection by the external flow to colloid diffusion over the yarn radius.
As the exterior flow strengthens, the colloids experience a modest advection-induced enhancement in withdrawal. Strikingly, in both S-O and S-I modes, larger flows sweep the solute rapidly from the domain, sharpening the solute gradients and thereby enhancing diffusiophoresis (Fig.~\ref{Fig:DP_performance_measure}(f)). At later times ($t\gg t_d$), these gradients decay, and colloid diffusion governs withdrawal irrespective of ambient flow strength as signaled by the late-time tails of profiles in Fig.~\ref{Fig:DP_performance_measure}(e).

\begin{figure}[H]
    \centering
\includegraphics[width=1\textwidth]{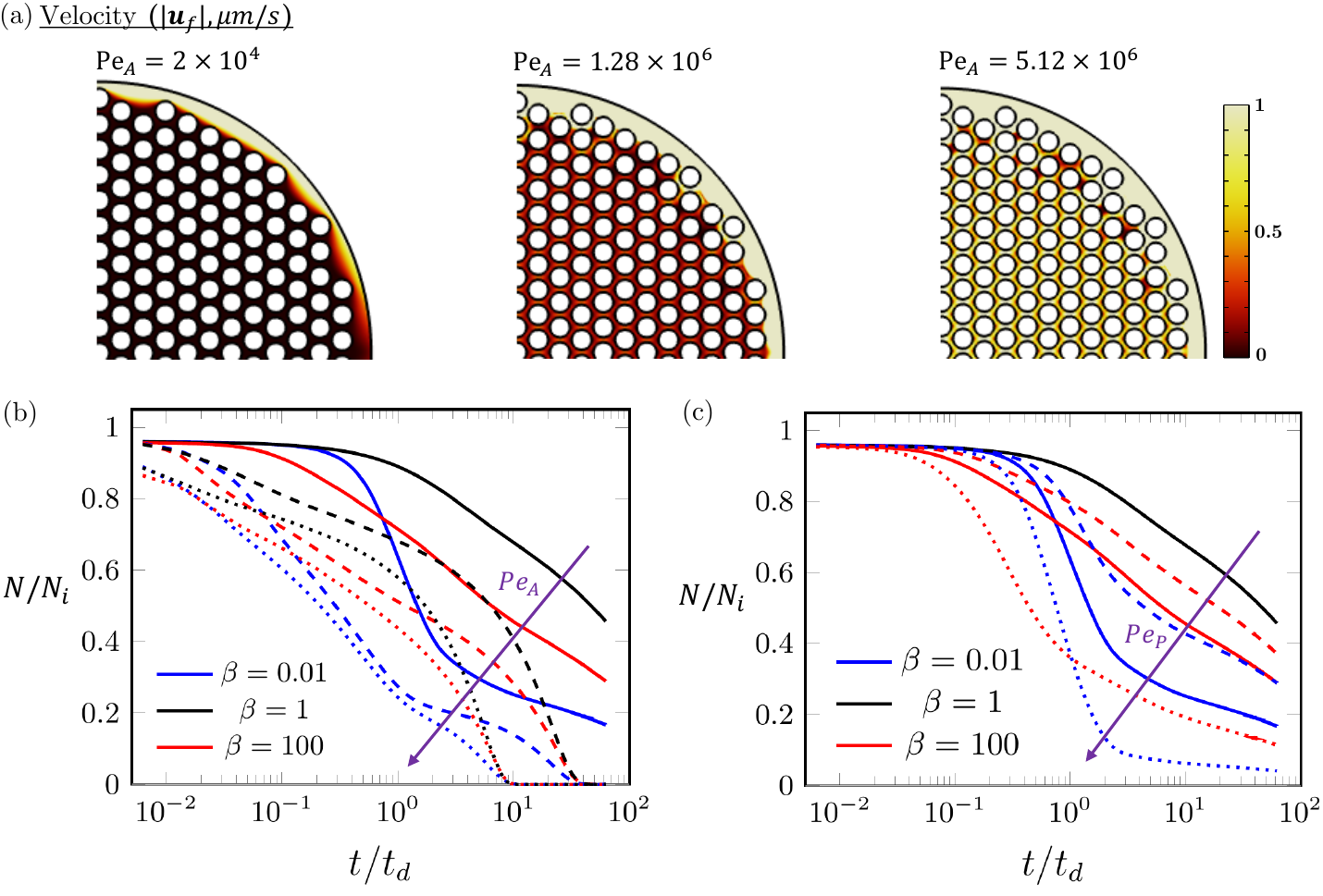}
    \caption{\small \textbf{Influence of $Pe_A$ and $Pe_P$ on colloid extraction.} (a) Velocity profiles for different $Pe_A$. (b) Variation of total number of colloids (per unit length) with time for different ambient Péclet numbers $Pe_A=2\times 10^4$ (solid lines), $1.28\times 10^6$ (dashed lines), $5.12\times 10^6$ (dotted lines). (c) Variation of total number of colloids (per unit length) with time for different colloid Péclet numbers ($Pe_P$). Here, solid lines, dashed lines, and dotted lines represent $\Gamma_p=\pm0.8\times 10^{-10}\,\mathrm{m}^2/\mathrm{s}$, $\pm2\times 10^{-10}\,\mathrm{m}^2/\mathrm{s}$, and $\pm5\times 10^{-10}\,\mathrm{m}^2/\mathrm{s}$, respectively. Parameters: $\mathcal{D}_s=10^{-9}\,\mathrm{m}^2/\mathrm{s}$ and $\mathcal{D}_p=10^{-12}\,\mathrm{m}^2/\mathrm{s}$.
 }
    \label{Fig:Influence of Pe_inf}
\end{figure}

To expand the parameter space, we now discuss breakthrough curves for various $Pe_A$, the phoretic Péclet number $Pe_P$ and gradient strength $\beta$.
Contour plots in Fig. \ref{Fig:Influence of Pe_inf}(a) show how an intense ambient flow penetrates the yarn and sweeps the internal stagnant zone. Fig. \ref{Fig:Influence of Pe_inf}(b) quantifies this for the no-gradient, S-O and S-I modes. Despite the strong convective sweep of solute through the core, both phoretic modes continue to offer better extraction over the examined range. These results reveal an existence of upper bound on the external flow strength up to which diffusiophoresis remains beneficial. 
However, this threshold lies in range outside the one pertinent to textile laundering.
Fig. \ref{Fig:Influence of Pe_inf}(c) shows the impact of phoretic Péclet number: as the mobility increases, extraction is expedited in both modes.

\subsection{Combined phoretic and osmotic extraction of colloids from porous media} 
\begin{figure*}[htbp]
    \centering
    \includegraphics[width=17.5cm]{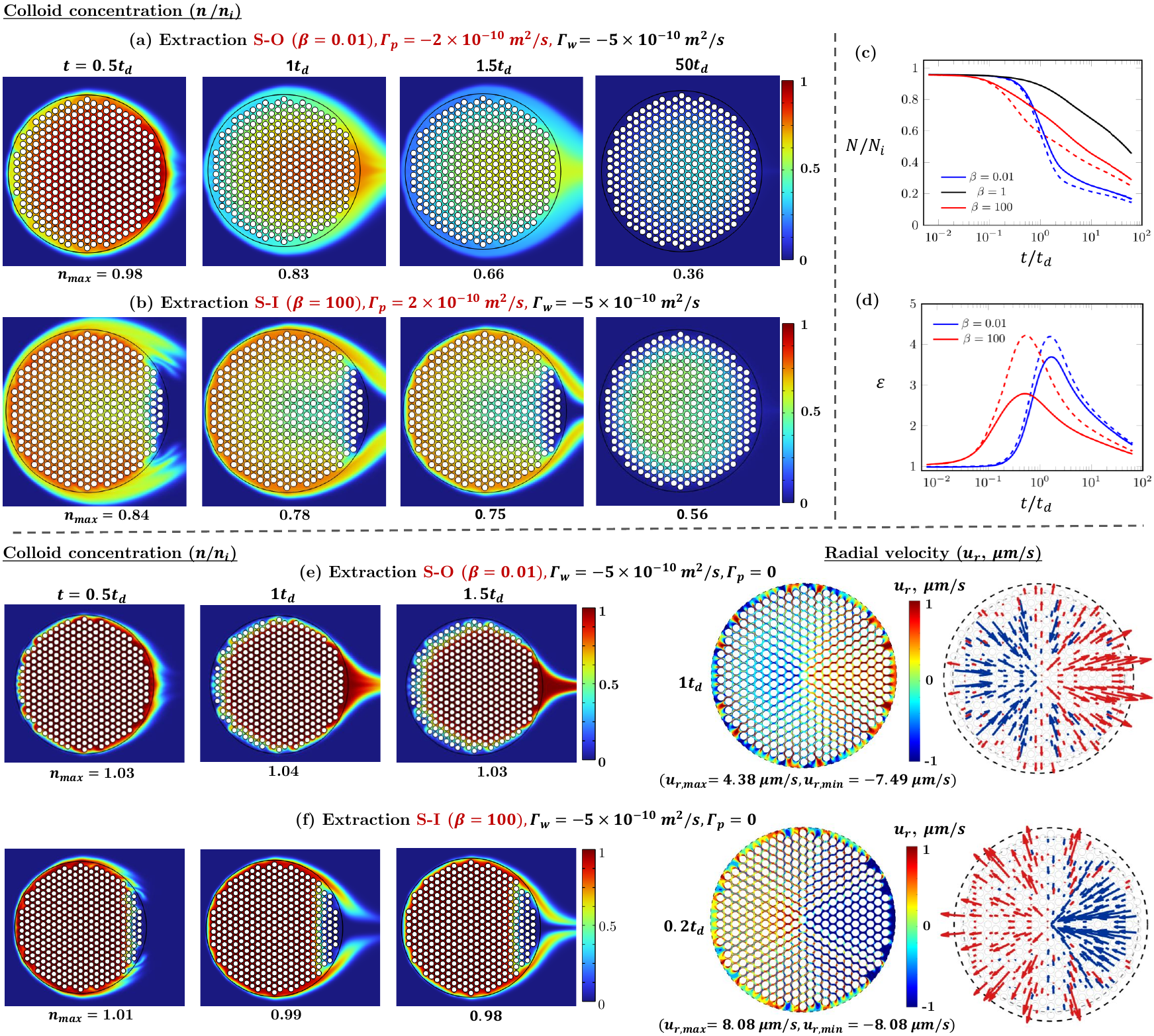} 
    \caption{\small \textbf{Influence of internal diffusio-osmosis on phoretic extraction.} 
    Spatiotemporal patterns of (a) solute-out ($\beta=0.01$) and (b) solute-in ($\beta=100$) modes for $\Gamma_w<0$. (c) Total number of colloids (per unit length). (d) Temporal evolution of effectiveness ($\varepsilon$) of colloid extraction with and without osmotic effects.
    (e,f) Spatiotemporal patterns of colloid extraction via pure osmosis in (e) solute-out and (f) solute-in mode. Figures on the right in (e,f) depict radially inward (blue) and outward (red) osmotic flow ($\mathbf{u}_r=(x\mathbf{u}_{fx}+y\mathbf{u}_{fy})/\sqrt{x^2+y^2}$) generating within the domain.
    Parameters: $\mathcal{D}_s = 10^{-9}\,\mathrm{m}^2/\mathrm{s}$, $\mathcal{D}_p = 10^{-12}\,\mathrm{m}^2/\mathrm{s}$, $\Gamma_p = \pm 2\times 10^{-10}\,\mathrm{m}^2/\mathrm{s}$, $\Gamma_w = - 5\times 10^{-10}\,\mathrm{m}^2/\mathrm{s}$.
    }
    \label{Fig:DP+DO_profile_-Gw}
\end{figure*}
\noindent Similar to colloids, surfaces of the porous domain can also exhibit significant mobility.
This slip velocity is accounted as $\mathbf{u}_w =-\Gamma_w \nabla \ln c $. 
Here $\Gamma_w$ is a negative wall mobility of porous media or fibre walls, and it is identical to Eq.(\ref{eq:GammaP}), where the zeta potential is $\zeta_w$.
\fig\ref{Fig:DP+DO_profile_-Gw} (a,b) shows how the internal osmotic flow  affects the phoretic extraction.
Spatiotemporal profiles depict that extraction in both modes is expedited, which is quantified by dashed curves in \fig\ref{Fig:DP+DO_profile_-Gw} (c). The enhancement profiles in \fig\ref{Fig:DP+DO_profile_-Gw} (d) depict a substantial enhancement in the time span until solute persists in the domain ($t \leq t_d$).
Interestingly, the addition of osmotic flow in S-I mode offers larger enhancement than S-O mode.

To understand this asymmetric enhancement, we explore the osmotic flow in isolation in \fig\ref{Fig:DP+DO_profile_-Gw}(e-f) i.e., $\Gamma_p=0$. 
The direction of the flow is illustrated by the velocity contours and vector plots, where red arrows denote radially outward flow and blue arrows denote inward flow.
We observe that osmotic flow in S-O mode releases fluid (containing colloids) from a narrower downstream region and draws in fluid from the remaining region, whereas S-I mode releases fluid from all over its periphery [note the peripheral band of colloids in \fig\ref{Fig:DP+DO_profile_-Gw}(f)] except this narrow downstream region.
However, the solute gradients are set in such a way that, for S-O  mode established under the current conditions $\Gamma_w < 0$, the osmotic slip should drive an inward flow across the entire domain. 
We do not observe this because, due to the implicit no-net-flow-rate constraint inside the stagnant yarn, incompressibility dictates that fluid must be expelled through the region of minimal osmotic pumping---specifically, the downstream region where the concentration gradient is weakest [see Fig.~\ref{Fig:DP_profiles}(c)] and is thus dominated by other regions, which results in a reversed flow. 
Thus, in S-O mode, a larger region experiences an inward osmotic velocity that counteracts the phoretic withdrawal. 
Conversely, the S-I mode follows the opposite scenario where the osmotic flow aids in withdrawal of particles over a relatively larger area.
Hence, this asymmetry in spatial orientation of osmotic flow in the two modes is responsible for a wider expulsion of particles in S-I mode and results in a larger enhancement boost in \fig\ref{Fig:DP+DO_profile_-Gw}(c,d).
Qualitatively similar results were observed for the positive wall mobility case and are detailed in the supplementary material. 
Further, we conducted a sensitivity analysis in terms of $\Lambda=\Gamma_w/\Gamma_p$. Increasing the $|\Lambda|$ modestly enhances colloid extraction by strengthening the osmotic advection inside the yarn (see \S6 in Supplementary material).

\section{Discussion and Outlook}
\label{sec:iv}

Colloidal extraction from porous media is limited by stagnant, dead-end regions inaccessible to bulk advection, a bottleneck directly relevant to enhanced oil recovery and textile laundering. 
By carrying out finite-element simulations, we show that diffusiophoretic (DP) and diffusioosmotic (DO) transport can rapidly mobilize colloids from these stagnant regions. The orientation of the applied solute gradient, solute-out (S-O) vs. solute-in (S-I), governs both the spatial pattern and overall efficiency of this extraction.
DP extraction is qualitatively distinct between the two modes: S-O withdraws colloids first from the periphery and then the core, while S-I withdraws them from the stagnant core outward. This spatial asymmetry has a temporal counterpart: S-I mobilizes colloids faster at early times, but S-O extracts more colloid over the phoretic lifespan. We also find that the persistence time in S-O grows logarithmically with the applied gradient strength, whereas in S-I it decays with gradient strength and becomes independent of it.

Since the confining walls of the porous structure can also exhibit mobility {\citep{kar2015enhanced,alessio2022diffusioosmosis}}, we next examine how the resulting internal osmotic flow affects colloidal extraction.
The osmotic slip at the porous walls of stagnant core enhances extraction in both modes, but the underlying flow pattern is reversed between them. In S-I mode, the osmotic slip drives an outward flow over nearly the entire domain, opposed only by a narrow downstream arc (of weakest solute gradient) where incompressibility constraint forces the flow to reverse. 
Conversely, in S-O mode, the osmotic slip is inward everywhere except a narrow downstream arc. 
Since the osmotic contribution expels fluid (containing colloids) outward over a much larger area in S-I mode, it reinforces DP-induced expulsion more strongly and produces a larger relative enhancement.
This asymmetric enhancement reflects that gradient orientation, in addition to wall mobility, governs how effectively osmotic slip can be harnessed to extract colloids.

These insights on gradient orientation-dependence and internal osmotic flow patterns can be leveraged for better colloid delivery in constricted pores \citep{rasmussen2020size}, enhanced oil recovery \citep{park2021microfluidic}, and laundry detergency \citep{shin2018cleaning}, where switching or combining gradient direction over a cycle could in principle deliver both rapid initial mobilization and extraction.
Building on this concept, cyclic switching of the gradient orientation through alternating S-I and S-O modes can periodically replenish solute within the pore space and modestly enhance phoretic colloid extraction (see Appendix B).
Our results further show that structural heterogeneity in the geometry, arising from disordered packing, does not qualitatively alter the influence of gradient orientation on colloid extraction (see Appendix C).
Similarly, the influence of porosity variation by changing the surface-to-surface gap between fibres ($S_f$) also does not affect the qualitative nature of our results (see \S 5 in Supplementary material).

\begin{figure*}[t!]
    \centering
\includegraphics[width=1 \textwidth]{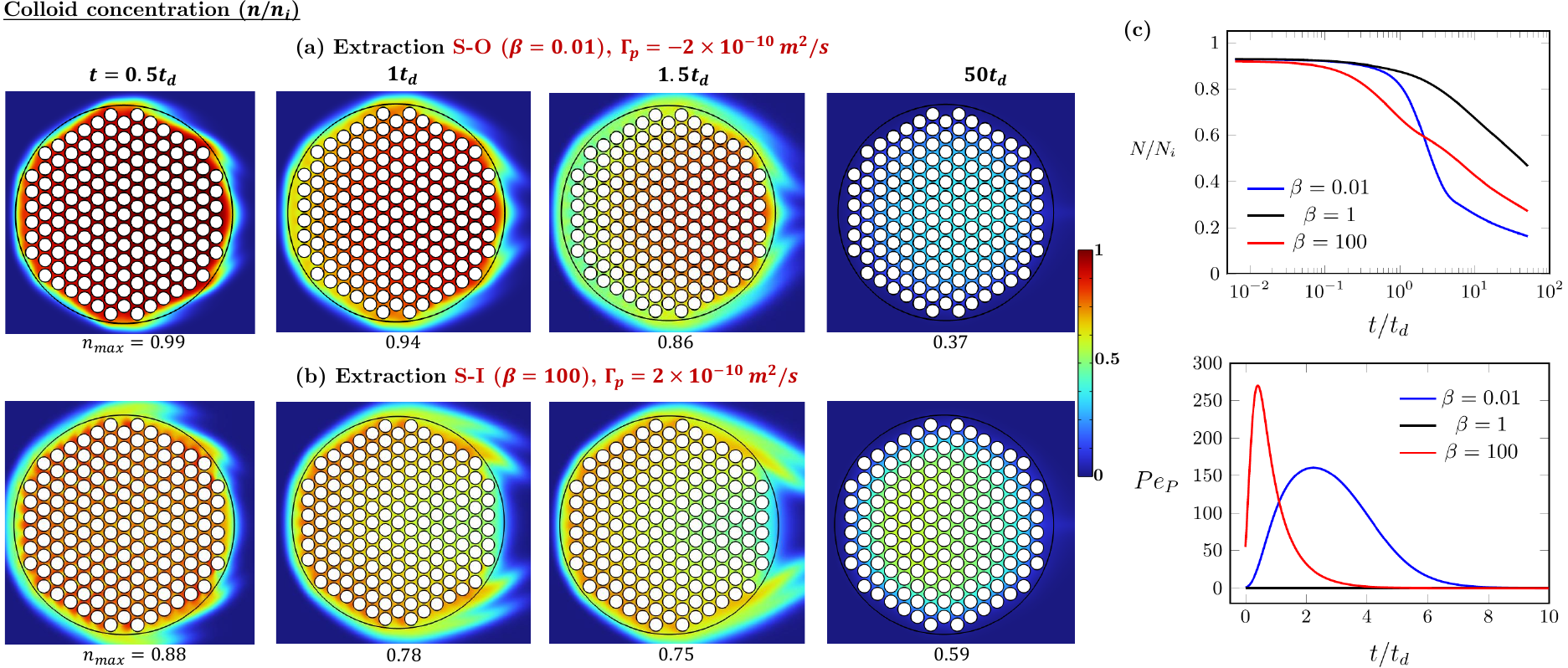}
    \caption{\small \textbf{Colloid profiles inside porous media.} Colloid concentration variation with time for (a) S-O ($\beta=0.01$) and (b) S-I ($\beta=100$) modes. (c) Total number of colloids (per unit length) variation with time for no-gradient control ($\beta=1$), S-O ($\beta=0.01$) and S-I ($\beta=100$) modes. Parameters: $\mathcal{D}_s=10^{-9}\,\mathrm{m}^2/\mathrm{s}$, $\mathcal{D}_p=10^{-12}\,\mathrm{m}^2/\mathrm{s}$, $\Gamma_p=\pm 2\times 10^{-10}\,\mathrm{m}^2/\mathrm{s}$, $R=250$ $\mu$m (yarn radius), $R_f=7.5$ $\mu$m (fibre radius), $S_f=2.5$ $\mu$m (spacing), $\phi=0.414$ (porosity), and $N_f=163$ (number of fibres).
    }
    \label{Fig:Realstic yarn_colloid_profiles}
\end{figure*}
Textile structures are intrinsically multi-porous, with transport occurring both between and within yarns \citep{warmoeskerken2002laundry}. 
Measurements report yarn diameters ranging from approximately $0.2$ to $0.7~\mathrm{mm}$, depending on the textile construction, with fiber diameters of order $10$--$20~\mu\mathrm{m}$ \citep{van1987hydrodynamics,puszkarz2020analysis}. The present choice of $R=400~\mu\mathrm{m}$ (yarn diameter $0.8~\mathrm{mm}$) therefore represents a relatively coarse but physically plausible yarn-scale structure. 
Additional simulations with smaller yarn radii (Fig. \ref{Fig:Realstic yarn_colloid_profiles}) exhibit the same qualitative behavior and trends.
The porosity is set to $\phi=0.48$, comparable to the $\phi\simeq0.45$ measured for monoporous polyester by \citet{van1987hydrodynamics}. For the fiber scale, we use $R_f=15~\mu\mathrm{m}$, corresponding to a fiber diameter of $30~\mu\mathrm{m}$, as a coarse representation of the fiber scale; reported fiber diameters are of order $10$--$20~\mu\mathrm{m}$ \citep{van1987hydrodynamics,puszkarz2020analysis}.

\begin{figure}[htbp]
    \centering
\includegraphics[width=1\textwidth]{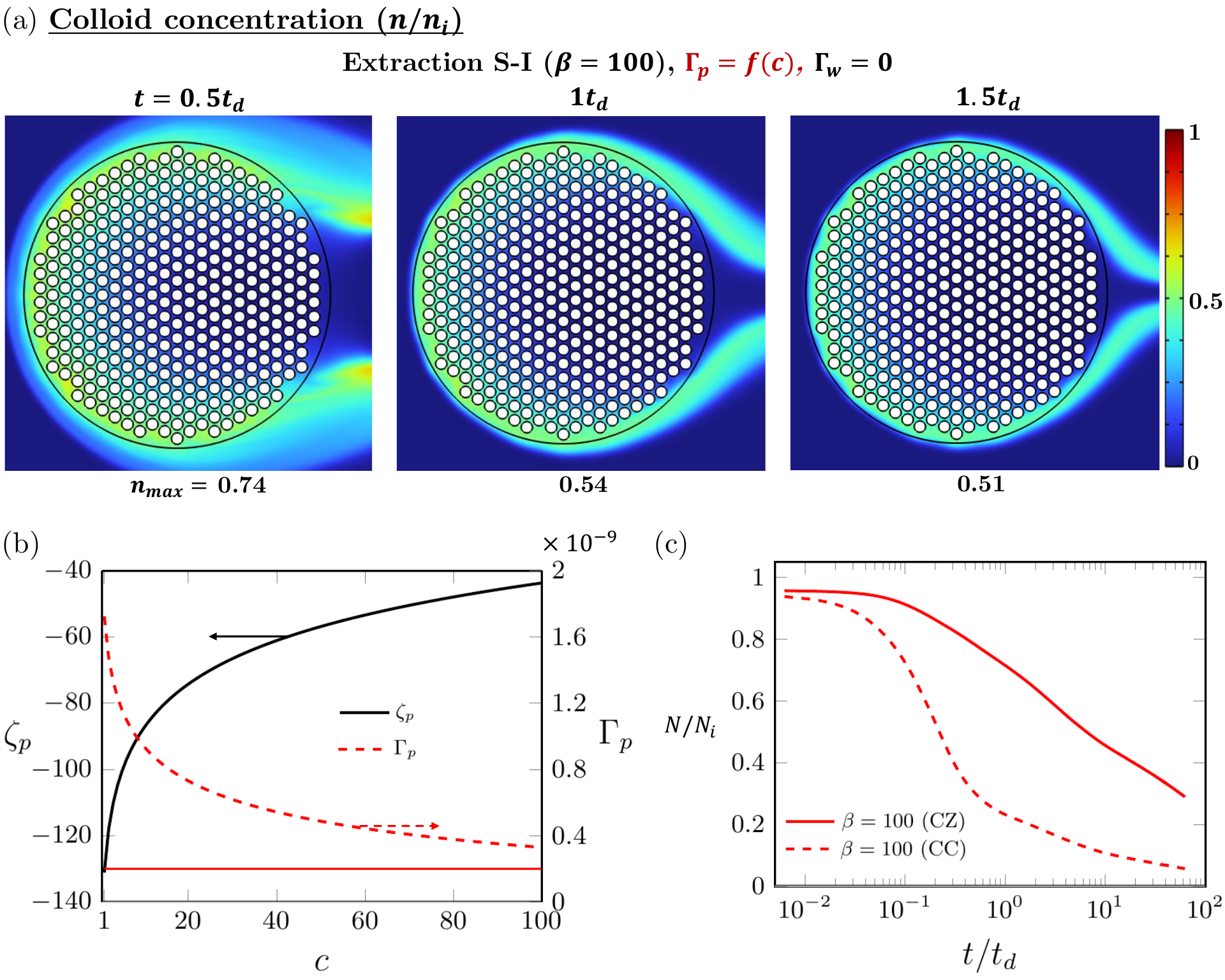}
    \caption{\small \textbf{Constant charge model.} (a) Colloid concentration variation with time for concentration-dependent mobility ($\Gamma_p=f(c)$). (b) Zeta potential ($\zeta$, mV) and colloid mobility ($\Gamma_p$, $\,\mathrm{m}^2/\mathrm{s}$) variation with solute concentration ($c$, mM). (c) Variation of total number of colloids per unit length with time.  The red solid and dashed lines depict concentration-independent ($\Gamma_p=2\times 10^{-10}\,\mathrm{m}^2/\mathrm{s}$) and concentration-dependent ($\Gamma_p=f(c)$) colloid mobility, respectively. Parameters: $\mathcal{D}_s=10^{-9}\,\mathrm{m}^2/\mathrm{s}$, $\mathcal{D}_p=10^{-12}\,\mathrm{m}^2/\mathrm{s}$, $R=400$ $\mu$m (yarn radius), $R_f=15$ $\mu$m (fiber radius), $S_f=7.5$ $\mu$m (spacing), and $\phi=0.48$ (porosity).
 }
    \label{Fig:Charge-dependent model}
\end{figure}
We extend our study by relaxing the assumption of constant colloid mobility and allowing the zeta potential to depend on the local solute concentration. 
This dependence gives rise to spatiotemporal variations in mobility, as changes in solute concentration modify the counter-ion shielding of surface charges and, consequently, the thickness of the electrical double layer. 
The expression for the zeta potential is motivated by the correlation proposed by \citet{kirby2004zeta} for a 1:1 electrolytes such as NaCl and KCl, given by $\zeta=-a_1\log c$, where $a_1=-43.7$ mV and $c$ (mM) denotes the solute concentration. 
This approximation holds for concentrations from approximately 1~$\mu$M to 100~mM~\citep{lee2023role}, which spans the range $c_i$ to $\beta c_i$ considered here.
The chosen expression provides a negative zeta potential, which yields positive colloid mobility in the NaCl solution.
Hence, S-I mode is only compatible for colloid extraction. Therefore, we focus on the S-I extraction case here. 
As Fig. \ref{Fig:Charge-dependent model}(a) shows, differences from the constant mobility model [refer Fig. 4(b)] are simply quantitative. The aforementioned conclusions are further strengthened due to large mobility variations because, as Fig. \ref{Fig:Charge-dependent model}(b) shows, zeta potential rises significantly at lower concentrations. 
Fig. \ref{Fig:Charge-dependent model}(c) shows the quantitative comparison of the total number of colloids (per unit length) for two models.

We also note that the present framework is a two-dimensional idealization, valid when the ambient flow is weakly non-uniform along the fiber axis. 
For gentle flows, the relevant parameter is $G=\dot{\gamma}R/U_{\infty}\sim R/L_{\mathrm{ext}}$, where $\dot{\gamma}$ is the shear rate of the exterior flow and $L_{\mathrm{ext}}$ the spacing between yarn crossover points in the weave. 
Since $G$ is one half of the fractional cover of a single thread set, it is governed by the tightness of the weave rather than the yarn size, and remains below $0.3$ for typical plain cotton woven fabrics \citep{ozgen2011yarn,zhu2018prediction}. 
Because phoretic and osmotic transport are both driven by $\nabla\ln c$, the axial drift should be weaker than the radial drift by $O(G)$. 
Hence the axial transport remains small compared with the yarn radius. Outside this regime ($\dot{\gamma} \gtrsim U_\infty/R$) longitudinal advection would compete with the radial drift, not currently captured in the current work.

Finally, we note that the present model treats the fiber walls as impermeable to
colloids, neglecting short-range particle-surface (DLVO) interactions and the associated deposition. 
Such effects can be incorporated without resolving the nanometer-scale interaction zone by following {\citet{spielman1974role}}, who showed that a thin interaction layer is equivalent to a first-order surface reaction, so that the DLVO potential can be lumped into a finite deposition rate constant $k_{\mathrm{dep}}$ imposed through a Robin condition, $-\mathcal{D}_p\,\nabla n\cdot\hat{\mathbf{n}} = k_{\mathrm{dep}}\,n$, at the walls. 
This caveat is chiefly for oppositely charged particles and
fibers (the combination already noted in \S II as experimentally
challenging) whereas like-charged surfaces retain a repulsive barrier that suppresses $k_{\mathrm{dep}}$ and leaves the current zero-flux idealization a good approximation. 
Since $k_{\mathrm{dep}}$ depends on the local salt concentration
and would therefore evolve differently in the two modes, quantifying its effect warrants a dedicated study.

\appendix

\section*{Appendix A: Mesh independence}
\label{sec:mesh_test}
\begin{figure}[H]
    \centering
    \includegraphics[width=1 \textwidth]{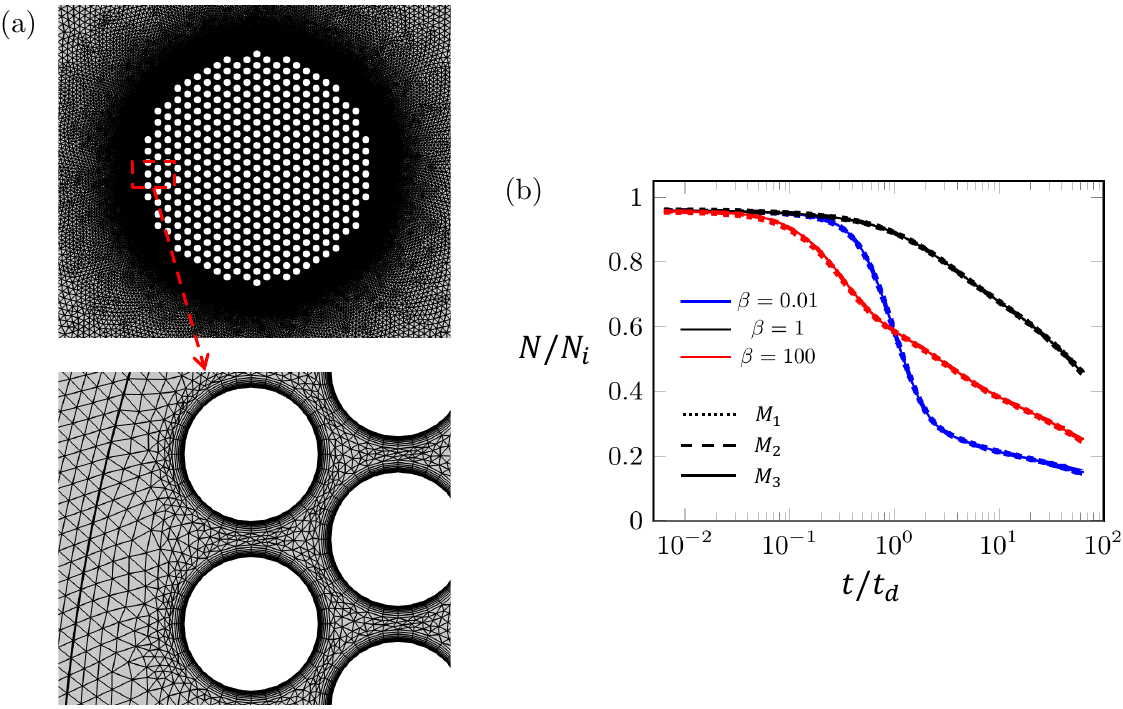}
    \caption{\small (a) Computational domain: yarn ($R=400$ $\mu$m) made up of bundle of fibers ($R_f=15$ $\mu$m). The inset shows an enlarged portion of domain discretized via a free triangular mesh. (b) Total number of colloids (per unit length) variation with time for various mesh elements. Parameters are identical to Fig.\ref{Fig:DP+DO_profile_-Gw}.
    }
    \label{Fig:mesh_test}
\end{figure}

\noindent
The computational domain is discretized using an unstructured, free triangular mesh with boundary layers on confined pore/fiber walls, as shown in \fig\ref{Fig:mesh_test} (a). A mesh independence test is carried out for the meshes: $M_1$, $ M_2$, and $ M_3$, consisting of $208312$ (degree of freedom DoF $= 783551$), $236466$ (DoF $= 854346$), and $295606$ (DoF $= 1002926$) triangular elements, respectively. The relative tolerance limit is set as $10^{-5}$. The results of the total number of colloids per unit length for S-O ($\beta=0.01$), no-gradient ($\beta=1$), and S-I ($\beta=100$) modes have shown insignificant changes (i.e., $\pm1-2 \%$) for considering mesh elements ($M_1$, $M_2$, and $M_3$) (refer \fig\ref{Fig:mesh_test} (b)). Hence, the mesh $M_3$ is used in this study for the final results.

\section*{Appendix B: Phoretic colloid extraction via cyclic salt gradient method}
\label{sec:cyclic salt gradient}
\begin{figure}[h]
    \centering
\includegraphics[width=1\textwidth]{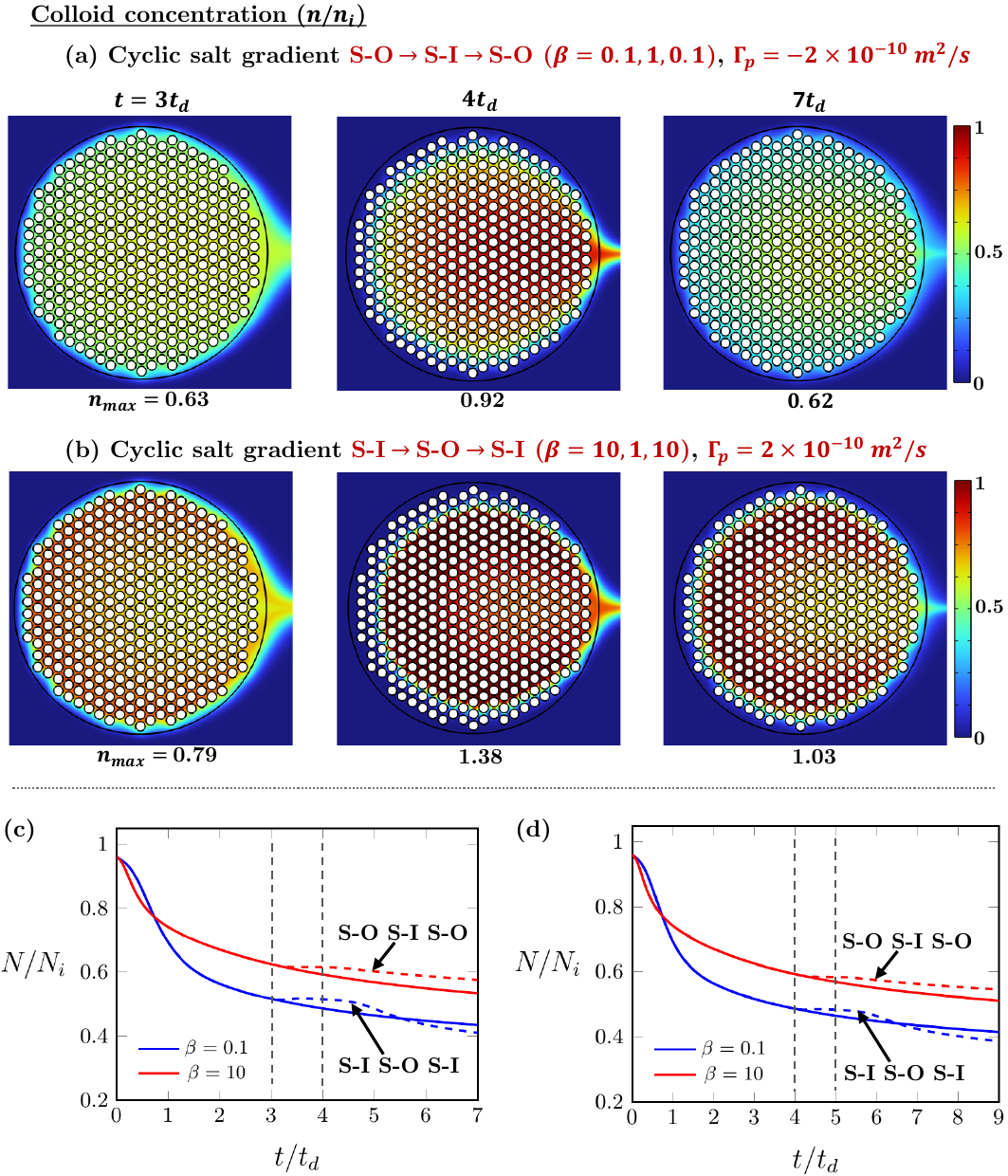}
    \caption{
    \small \textbf{Colloid profile in porous media for cyclic salt gradients.} Colloid concentration variation with time for (a) S-O $\rightarrow$ S-I $\rightarrow$ S-O and (b) S-I $\rightarrow$ S-O $\rightarrow$ S-I cyclic ($3t_d$-extraction--$1\,t_d$-replenish) mode. Colloid concentration evolution with time for (c) $3-1-3$ and (d) $4-1-4$ cyclic modes. Here, solid lines represent the single-gradient mode; dashed lines denote the cyclic modes. Parameters: solute ambipolar diffusivity $\mathcal{D}_s=10^{-9}$ $m^2 /s$, colloid diffusivity $\mathcal{D}_p=10^{-12}$ $m^2 /s$, colloid mobility $\Gamma_p=\pm 2\times 10^{-10}$ $m^2/s$, yarn radius $R=400$ $\mu m$, fiber radius $R_f=15$ $\mu m$, surface-to-surface gap between fibers $S_f=7.5$ $\mu m$, and porosity $\phi=0.48$. 
 }
    \label{Fig:Modes cycle profiles}
\end{figure}

\noindent
In textile laundering, cyclic switching can be implemented when automated machines alternate between detergent-dosing (solute-in) and fresh-water rinsing (solute-out) modes.
Simulations in Fig. \ref{Fig:Modes cycle profiles}(a,b) show that cyclically switching $\beta$ by altering the external flow composition periodically replenishes the local pore space with solute and thus can lead to modestly enhanced extraction. 

As dashed curve in Fig.\ref{Fig:Modes cycle profiles}(c) shows, after a diffusiophoretic extraction cycle of S-O mode until $\sim 3 t_d$, a S-I  cycle replenishes solute for $t_d$ time (thereby compacting the particles inside the domain), a subsequent S-O cycle can more efficiently extract particles. The choice of $3 t_d$ is motivated by the observation that diffusiophoresis is significantly persistent until those times (see Fig. 5). 

Interestingly, the converse cycle S-I $\to$ S-O $\to$ S-I in the $3t_d$-extraction--$1\,t_d$-replenish mode does not offer better extraction, because phoresis in S-O mode sustains longer. 
This is because, as the interior concentration falls during extraction, the logarithmic gradient $\nabla \ln c$ is progressively amplified, whereas the rising interior concentration in S-I mode attenuates it.
Fig. \ref{Fig:Modes cycle profiles}(d) shows qualitatively similar results for $4t_d$-extraction-$1_d$-replenish mode.
The slight rise in colloid concentration during the replenishment phase occurs because the colloids in the immediate vicinity of the yarn are driven back inside due to the compaction induced by the solute-gradient reversal. The magnitude of this increment diminishes as the switching interval increases from the 3-1-3 to the 4-1-4 mode.

\section*{Appendix C: Effect of structural heterogeneity on colloid extraction}
\label{sec:structural heterogeneity}

\begin{figure}[h]
    \centering
\includegraphics[width=1 \textwidth]{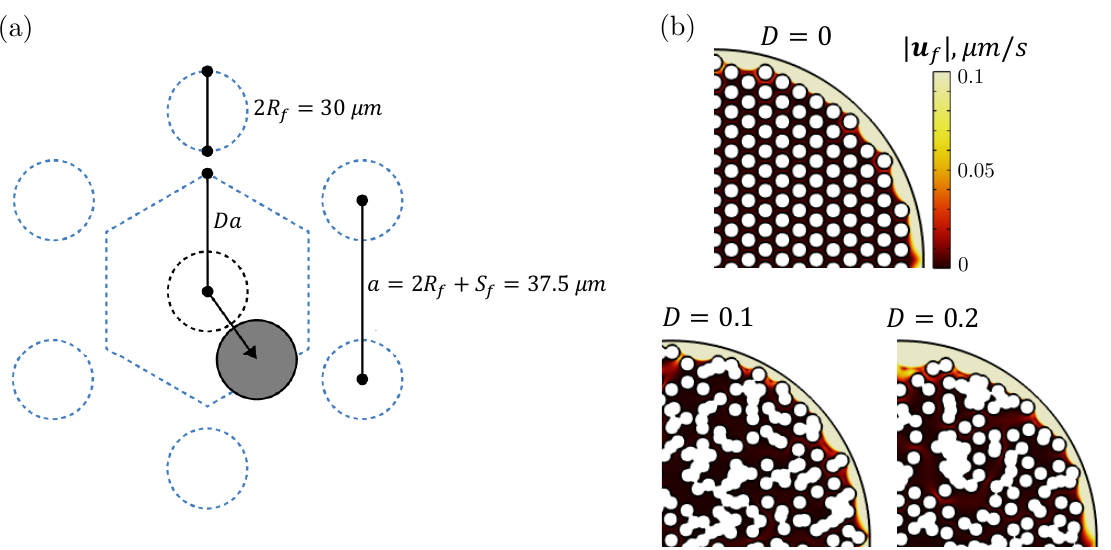}
    \caption{\small (a) Schematic of the disordering mechanism in fibre arrays. Disordered fibre arrays (radius, $R_f$) are created by randomly shifting the fibres of a regular hexagonal lattice (lattice constant, ($a =2R_f +S_f$)). Each fibre is randomly placed within a hexagon of circumradius ($Da$) \citep{walkama2020disorder,haward2021stagnation}. (b) Velocity ($|\mathbf{u}_f|, \mu m/s$) profiles in the ordered ($D=0$) and disordered ($D=0.1,0.2$) porous structures. Parameters: $R=400$ $\mu$m (yarn radius), $R_f=15$ $\mu$m (fibre radius), $S_f=7.5$ $\mu$m (spacing), and $\phi=0.48$ (porosity).
    }
    \label{Fig:Disordered_vel_and_Pec}
\end{figure}
\begin{figure*}[h]
    \centering
\includegraphics[width=1 \textwidth]{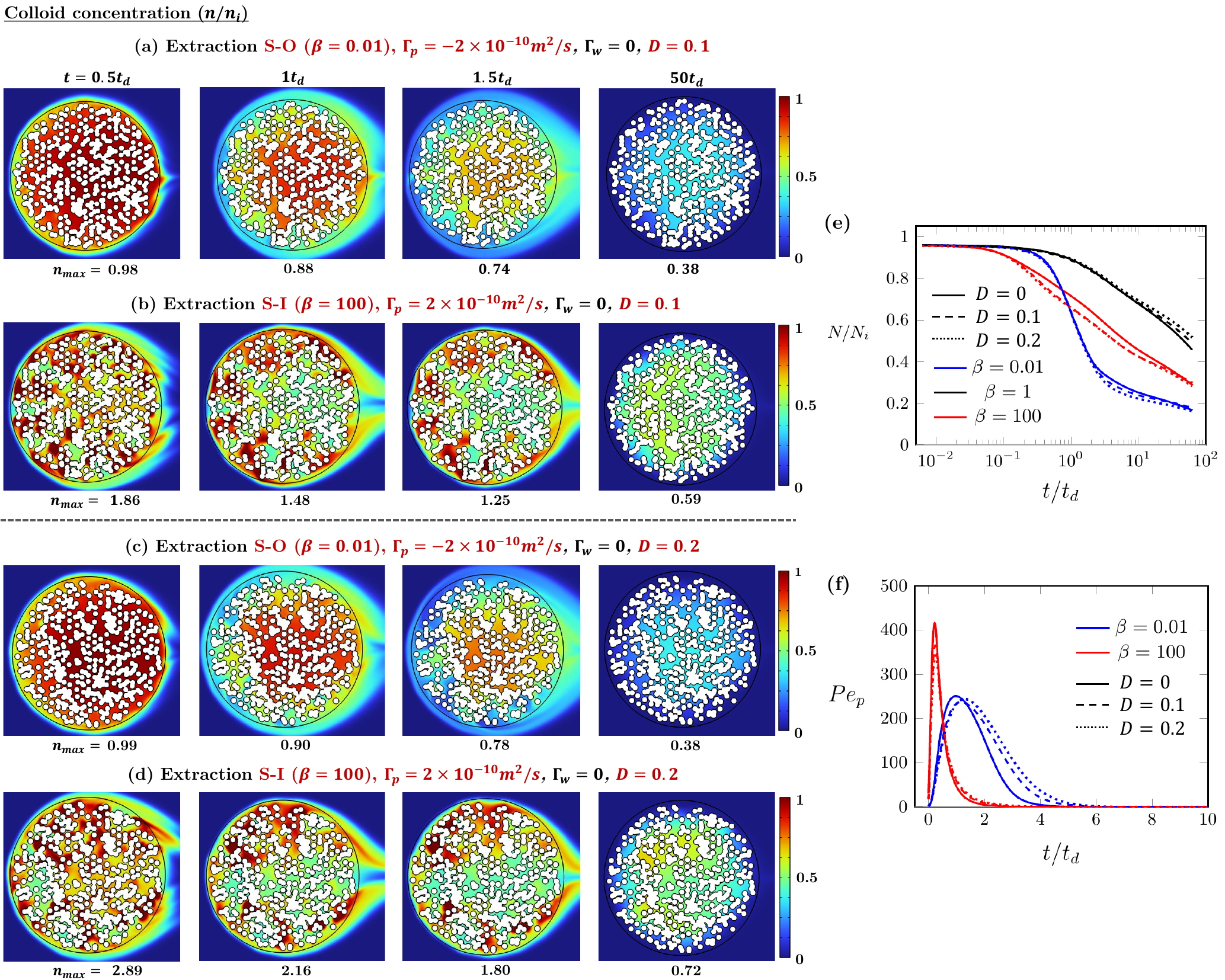}
    \caption{\small \textbf{Colloid profiles in disordered porous media.} Colloid concentration variation with time for (a,b) $D=0.1$ and (c,d) $D=0.2$ for both S-O ($\beta=0.01$) and S-I ($\beta=100$) modes. Variation of (e) total number of colloid (per unit length) and (f) phoretic Péclet number with time in ordered ($D=0$), $10\%$ disordered ($D=0.1$), and $20\%$ disordered ($D=0.2$) porous media. Parameters: $\mathcal{D}_s=10^{-9}\,\mathrm{m}^2/\mathrm{s}$, $\mathcal{D}_p=10^{-12}\,\mathrm{m}^2/\mathrm{s}$, $\Gamma_p=\pm 2\times 10^{-10}\,\mathrm{m}^2/\mathrm{s}$, $R=400$ $\mu$m (yarn radius), $R_f=15$ $\mu$m (fibre radius), and $\phi=0.48$ (porosity).
 }
    \label{Fig:Disordered_colloid_profiles}
\end{figure*}
Here, we have shown the effect of structural heterogeneity via disorder in packing. 
We model disorder by displacing each fiber from its hexagonal lattice site via a random offset (see Fig. \ref{Fig:Disordered_vel_and_Pec}(a)),
\begin{gather}
R_{hex} = D \left( 2R_f + S_f \right), \notag \\
(x_i, y_i) = \left(x_i^{ordered}, y_i^{ordered}\right) + (\Delta x_i, \Delta y_i)
\label{eq:Rhex}
\end{gather}
\noindent where, $x_i$ and $y_i$ represent the new fiber center position. The  $\Delta x_i$ and $\Delta y_i$ are random shifts.
For $D = 0$, the fibers remain an ordered hexagonal lattice. For $D > 0$, each fiber is randomly displaced by an amount proportional to $D$ (e.g., $D = 0.10$ corresponds to $10\%$ disorder, $D = 0.20$ corresponds to $20\%$ disorder). Increasing $D$ increases the degree of positional randomness in the fiber packing \citep{walkama2020disorder,haward2021stagnation}.
The porosity is held at $\phi = 0.48$ and only the pore-size distribution and connectivity vary. The flow intensity in ordered and disordered porous yarn is depicted in Fig. \ref{Fig:Disordered_vel_and_Pec}(b) which show the existence of stagnant core for all cases. 
Figs. \ref{Fig:Disordered_colloid_profiles}(a-d) shows that, despite the disorder, the spatial signatures of the two modes are qualitatively similar to that observed in ordered yarn. 
Solute-out extraction still proceeds inward from the periphery, and solute-in extraction still clears the stagnant core first. 

 Fig. \ref{Fig:Disordered_colloid_profiles}(e) quantifies this effect on net withdrawal in three modes. Without a gradient ($\beta = 1$), disorder slightly retards extraction at larger time scales, consistent with the expected increase in tortuosity. 
Under an applied concentration gradient, the trend reverses, and both SO \& SI modes extract marginally more colloid at late times.
Fig. \ref{Fig:Disordered_colloid_profiles}(f) identifies the origin of this reversal. Disorder lowers and delays the peak of the colloid P\'eclet number, from $Pe_P \approx 416$ to $342$ in solute-in mode. 
This is because the solute drains slightly more slowly from the poorly connected pockets created by disorder, hence the local logarithmic gradient is weaker at any instant, however it persists over longer times due to these pockets.

\begin{acknowledgments}
The authors gratefully acknowledge financial support from the Anusandhan National Research Foundation (ANRF), India, under grant ANRF/ECRG/2025/002557/ENS.
\end{acknowledgments}

\section*{Supporting Information}
Validation of the ambient flow field against two-dimensional flow past a cylinder (\S1); boundary conditions and mesh independence study (\S3); colloid compaction under the alternative gradient--mobility pairing (\S4); effect of porosity and fibre spacing on colloid extraction (\S5); sensitivity of extraction to the mobility ratio $\Lambda = \Gamma_w/\Gamma_p$ and results for positive wall mobility (\S6) 

\clearpage

\bibliography{references}

\clearpage
\onecolumngrid

\setcounter{section}{0}

\renewcommand{\thesection}{S\arabic{section}}

\setcounter{figure}{0}
\renewcommand{\thefigure}{S\arabic{figure}}

\setcounter{table}{0}
\renewcommand{\thetable}{S\arabic{table}}

\setcounter{equation}{0}
\renewcommand{\theequation}{S\arabic{equation}}

\section*{Supplementary Material}

\subsection*{1. Flow past a porous cylinder} 
 \noindent The stream function near the surface for a two-dimensional flow past a cylinder is defined as follows \citep{proudman1957expansions,batchelor2000introduction}.
 \begin{gather}
\psi (r,\theta) = -\frac{R U_\infty}{\ln (Re)} \sin\theta \left(\frac{r}{R} \ln\frac{r}{R} - \frac{1}{2} \frac{r}{R} + \frac{1}{2} \frac{R}{r} \right)
 \end{gather}
 where $R$, $U_\infty$, and $Re$ are the radius of the cylinder, ambient flow velocity, and Reynolds number. The stream function in the Cartesian coordinates is expressed as follows. 
\begin{gather}
\psi (x,y) = -\frac{U_\infty}{\ln (Re)} y \left[\ln\frac{\sqrt{x^2+y^2}}{R} - \frac{1}{2} + \frac{R^2}{2(x^2+y^2)} \right]
 \end{gather}
\begin{figure}[H]
    \centering
    \includegraphics[width=1 \textwidth]
    {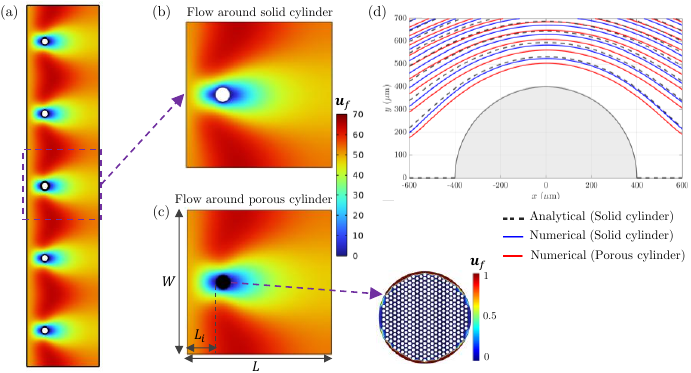}
    \caption{\small {Velocity ($\mathbf{u}_f$, $\mu m/s$) profiles for flow past (a) parallel array of solid cylinders, (b) single solid cylinder and (c) single porous cylinder. (d) Comparison of analytical and numerical stream function ($\psi$) near the surface for flow around a solid cylinder with numerical stream function for the present study (i.e., flow around a porous yarn). Parameters: $R=400 \mu m$, $U_\infty=50 \mu m/s$, and $Re=0.02$.}}
    \label{Fig:validation}
\end{figure}

 \clearpage

\subsection*{2. Combined phoretic and osmotic colloid extraction from porous media for $\Gamma_w>0$} 

\begin{figure*}[htbp]
\centering \includegraphics[width=17.5cm]{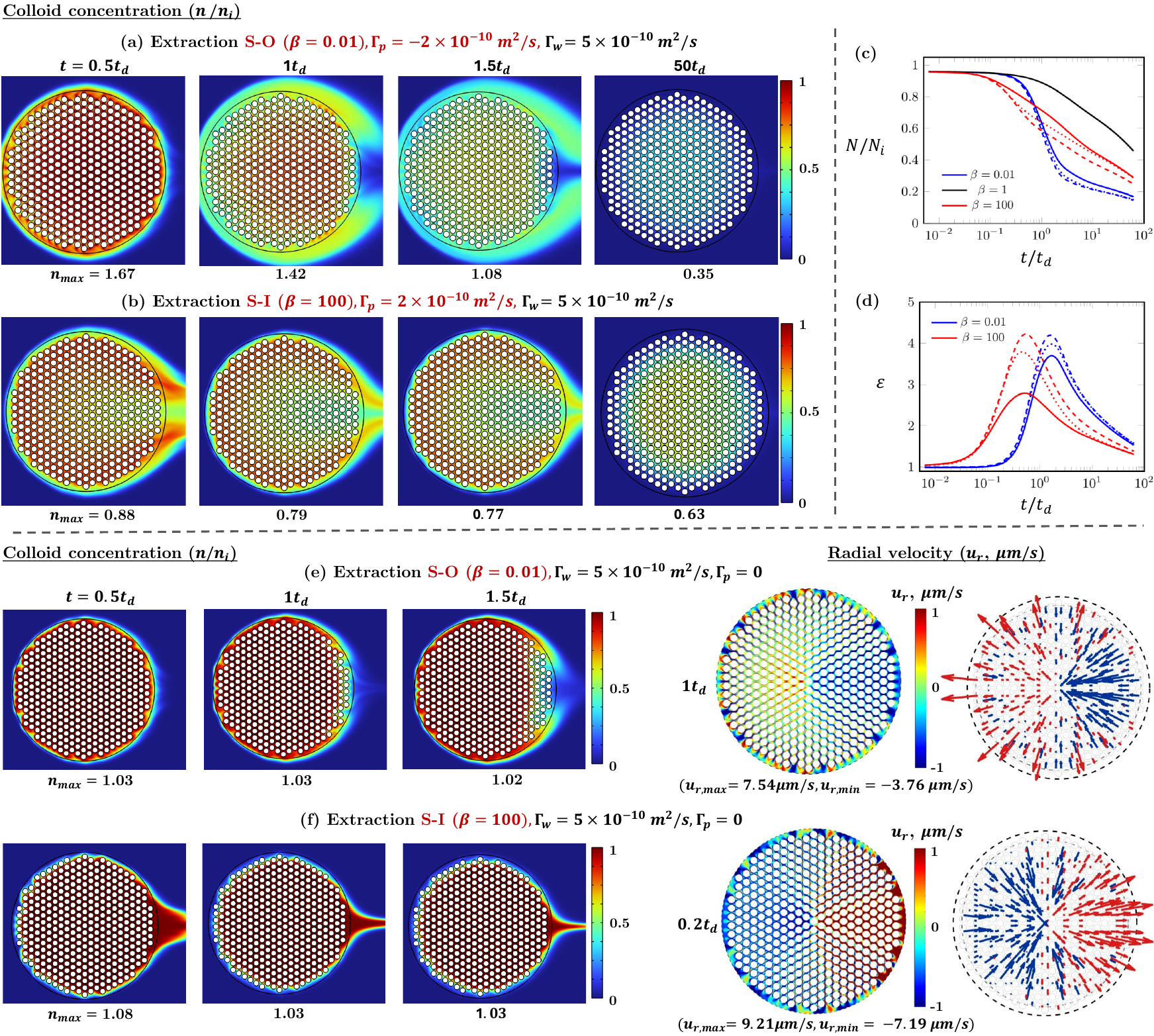}
    \caption{\small { \textbf{Influence of internal diffusio-osmosis on phoretic extraction.} 
    Spatiotemporal patterns of (a) solute-out ($\beta=0.01$) and (b) solute-in ($\beta=100$) modes for $\Gamma_w>0$. (c) Domain-averaged colloid concentration. (d) Temporal evolution of effectiveness ($\varepsilon$) of colloid extraction with and without osmotic effects. Here, solid, dashed and dotted lines denote $\Gamma_w=0$, $\Gamma_w<0$, and $\Gamma_w>0$, respectively. 
    (e,f) Spatiotemporal patterns of colloid extraction via pure osmosis in (e) solute-out and (f) solute-in mode. Figures on the right in (e,f) depict radially inward (blue) and outward (red) osmotic flow generating within the domain.
    Parameters: $\mathcal{D}_s = 10^{-9}\,\mathrm{m}^2/\mathrm{s}$, $\mathcal{D}_p = 10^{-12}\,\mathrm{m}^2/\mathrm{s}$, $\Gamma_p = \pm 2\times 10^{-10}\,\mathrm{m}^2/\mathrm{s}$, $\Gamma_w =  5\times 10^{-10}\,\mathrm{m}^2/\mathrm{s}$.}
    }
    \label{Fig:DP+DO_profile_Gw}
\end{figure*}

\clearpage

\subsection*{3. Boundary conditions and additional mesh independence study} 
\begin{figure}[H]
    \centering
\includegraphics[width=0.8\textwidth]{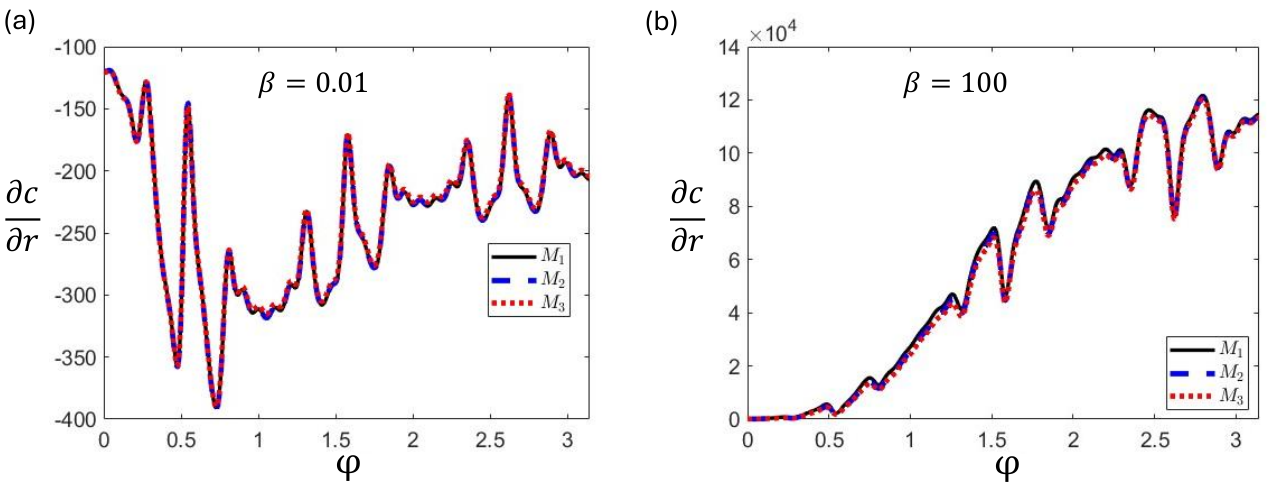}
    \caption{\small Solute concentration gradient ($\partial c/\partial r$, mol/m$^4$) variation with arc angle ($\varphi$, rad) for different mesh elements ($M_1,M_2$, and $M_3$) at $1t_d$ (for S-O) and $0.2t_d$ (for S-I). Parameters: solute ambipolar diffusivity $\mathcal{D}_s=10^{-9}$ $m^2 /s$ and mesh elements $M_1=208312$, $M_2=236466$, $M_3=295606$.
 }
    \label{Fig:solute_gradinet_mesh}
\end{figure}

\noindent
 The boundary conditions for the dynamics of the solute, colloid, and fluid are expressed as follows.
\begin{gather}
    \text{Initial solute and colloid concentration outside the porous structure} \quad c(x,y,0) = c_i \quad \text{and} \quad  n(x,y,0) = 0 \\
    \text{Initial solute and colloid concentration inside the porous structure} \quad c(x,y,0) = c_i \quad \text{and} \quad  n(x,y,0) = n_i \\
    \text{Fixed inlet solute and colloid concentration} \quad c(0,y,t) = \beta c_i \quad \text{and} \quad  n(0,y,t) = 0 \\
    \text{No-flux condition at outer domain walls, outlet, and fiber walls} \quad \nabla_n c = 0, \quad \nabla_n n = 0 \\
    \text{Fixed inlet average flow velocity} \quad \mathbf{u}_{fx}(0,y,t) = U_\infty \\
    \text{Fixed outlet pressure} \quad p(x,y,t) = 0 \\
    \text{Symmetry condition at outer domain walls} \quad \frac{\partial \mathbf{u}_{fx}}{\partial y} =0, \quad \mathbf{u}_{fy} = 0  \\
    \text{Diffusioosmotic wall slip condition on the fiber walls} \quad \textbf{u}_w = -\Gamma_w \nabla_t (\ln c)
\end{gather}
where $\nabla_n$ and $\nabla_t$ are the gradients normal and tangential to the boundary, respectively. The $c$, $n$, $c_i$, $n_i$, $\beta$, $\mathbf{u}_{fx}$, $\mathbf{u}_{fy}$, $U_\infty$, $p$, $\mathbf{u}_w$, $\Gamma_w$, $t$, $x$, and $y$ represent the solute and colloid concentration, initial concentration of solute and colloid, solute gradient strength, $x-$ and $y-$ component of velocity, ambient flow velocity, pressure, diffusioosmotic slip velocity, wall mobility, time, $x$ and $y$ coordinates, respectively.
\\
We have carried out a mesh independence test for the mesh elements $M_1$, $M_2$ and $M_3$ for local salt gradients (Fig. \ref{Fig:solute_gradinet_mesh}), peak phoretic velocity (Table \ref{tab:upmax}), and peak flow velocity (included both advection and osmotic velocity) (Table \ref{tab:umax}) inside the porous yarn. The results depicted an insignificant change in the quantities as the mesh elements varied from $M_1$ to $M_3$. 

\begin{table}[H]
\centering
\begin{tabular}{ccc}
\hline
 & \multicolumn{2}{c}{$\langle \mathbf{u}_p \rangle_{max}$} \\ \hline
Mesh & $\beta=0.01$ & $\beta=100$ \\ \hline
$M_1$ & 0.6341  & 1.0491\\ \hline
$M_2$ & 0.6362 & 1.0549 \\ \hline
$M_3$ & 0.6262 & 1.0527  \\ \hline
\end{tabular}
\caption{Maximum value of domain-averaged phoretic velocity ($\langle \mathbf{u}_p \rangle_{max}$, $\mu m/s$) variation with different mesh elements at $1t_d$ (for S-O) and $0.2t_d$ (for S-I). Parameters: solute ambipolar diffusivity $\mathcal{D}_s=10^{-9}$ $m^2 /s$, colloid mobility $\Gamma_p=\pm 2\times 10^{-10}$ $m^2 /s$, and mesh elements $M_1=208312$, $M_2=236466$, $M_3=295606$.}
\label{tab:upmax}
\end{table}
\begin{table}[H]
\centering
\begin{tabular}{ccc}
\hline
& \multicolumn{2}{c}{$\langle |\mathbf{u}_f| \rangle_{max}$} \\ \hline
Mesh & $\beta=0.01$ & $\beta=100$ \\ \hline
$M_1$ &  1.1797 & 1.6703 \\ \hline
$M_2$ & 1.1844 & 1.6846\\ \hline
$M_3$ & 1.1909 & 1.6804  \\ \hline
\end{tabular}
\caption{Maximum value of domain-averaged flow velocity ($\langle |\mathbf{u}_f| \rangle_{max}$, $\mu m/s$) variation with different mesh elements at $1t_d$ (for S-O) and $0.2t_d$ (for S-I). Parameters: solute ambipolar diffusivity $\mathcal{D}_s=10^{-9}$ $m^2 /s$, wall mobility $\Gamma_p=- 5\times 10^{-10}$ $m^2 /s$, and mesh elements $M_1=208312$, $M_2=236466$, $M_3=295606$.}
\label{tab:umax}
\end{table}

\subsection*{4. Colloid extraction-incompatible mobility-gradient combinations} 
\begin{figure}[H]
    \centering
\includegraphics[width=0.9\textwidth]{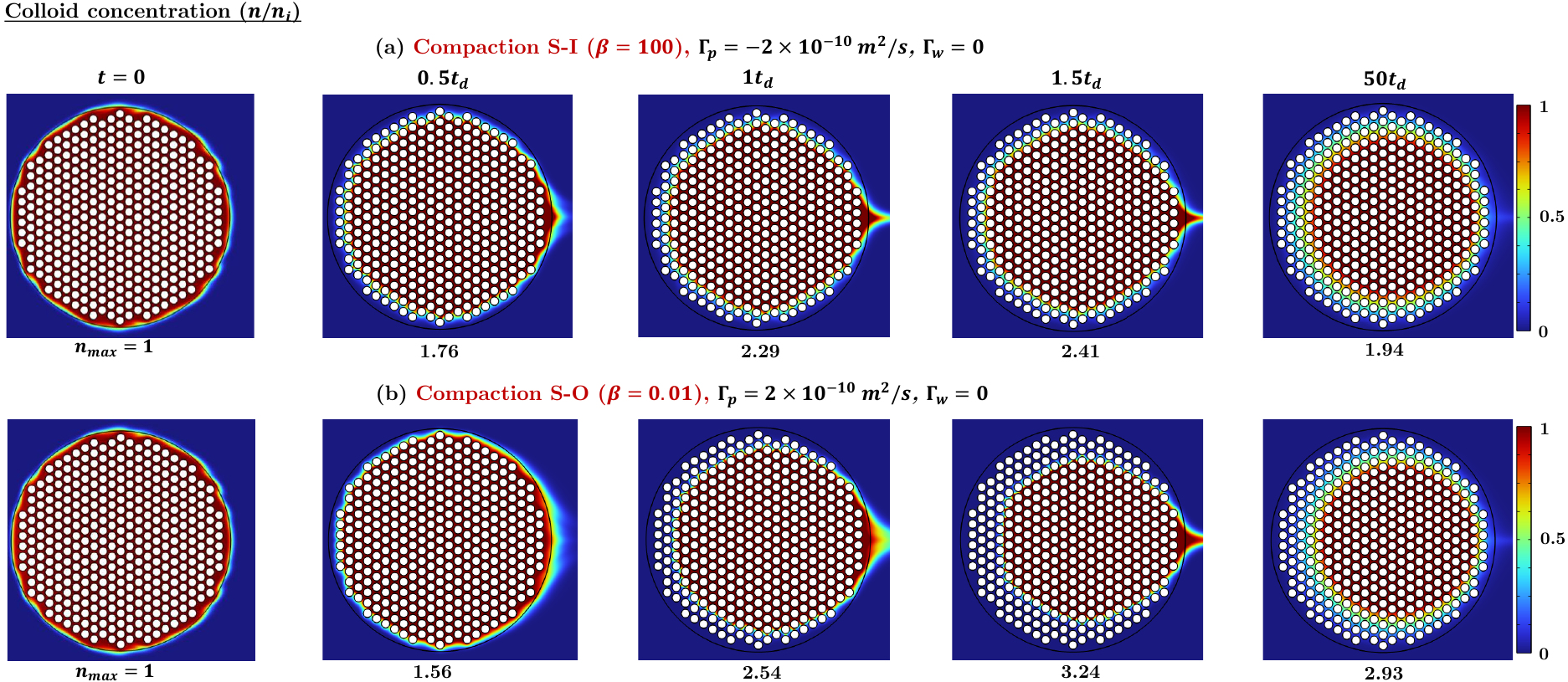}
    \caption{\small \textbf{Colloid profiles for extraction-incompatible mobility-gradient combinations.} Colloid concentration variation with time for (a) S-I compaction ($\beta=100, \Gamma_p<0$), (b) S-O compaction ($\beta=0.01, \Gamma_p>0$), respectively. Parameters: $\mathcal{D}_s=10^{-9}$ $m^2 /s$, $\mathcal{D}_p=10^{-12}$ $m^2 /s$, $\Gamma_p=\pm 2\times 10^{-10}$ $m^2/s$, $R=400$ $\mu m$ (domain radius), $R_f=15$ $\mu m$ (fiber radius), $S_f=7.5$ $\mu m$ (spacing), and $\phi=0.48$ (porosity).
 }
    \label{Fig:Extraction_and_compaction_colloid_profiles}
\end{figure}

\begin{figure}[H]
    \centering
\includegraphics[width=0.35 \textwidth]{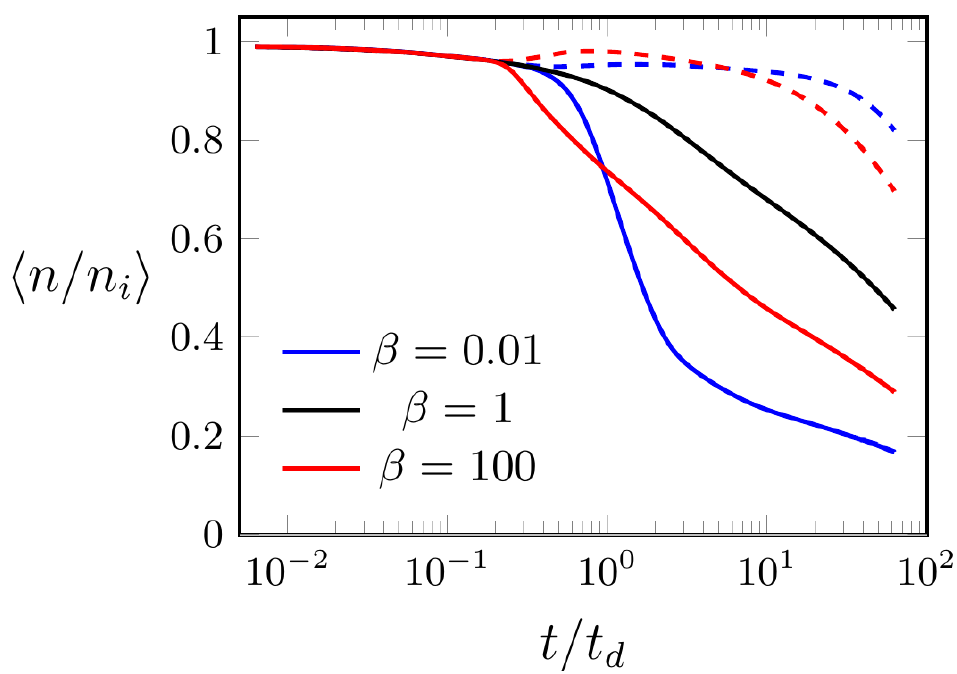}
    \caption{\small \textbf{Quantification of colloid extraction and compaction.} Solid and dashed lines depict the extraction and compaction of the colloid, respectively.
The slight rise in colloid concentration occurs because the colloids in the immediate vicinity of the yarn (convected at periphery) are driven back inside due to the compaction induced by the solute-gradient (competing with diffusion and convection at the periphery). Parameters: $\mathcal{D}_s=10^{-9}$ $m^2 /s$, $\mathcal{D}_p=10^{-12}$ $m^2 /s$, and $\Gamma_p=\pm 2\times 10^{-10}$ $m^2/s$.
 }
    \label{Fig:Extraction_and_compaction_colloid_quantification}
\end{figure}

\noindent 
In this section, we explore extraction-incompatible mobility-gradient combinations. In the main text, we have selected the ‘extraction-compatible’ combination of parameters. We did this because the other sets of zeta potentials results in inward \textbf{compaction} of trapped colloids because diffusiophoresis acts inward for opposite mobilities; this is shown in Figs. \ref{Fig:Extraction_and_compaction_colloid_profiles} and \ref{Fig:Extraction_and_compaction_colloid_quantification} for identical flow and geometry values. As the colloid concentration contours depict, injection is counter-productive to colloid extraction.

\subsection*{5. Influence of porosity ($\phi$) variation on phoretic colloid extraction} 
\begin{figure}[H]
    \centering
\includegraphics[width=0.9 \textwidth]{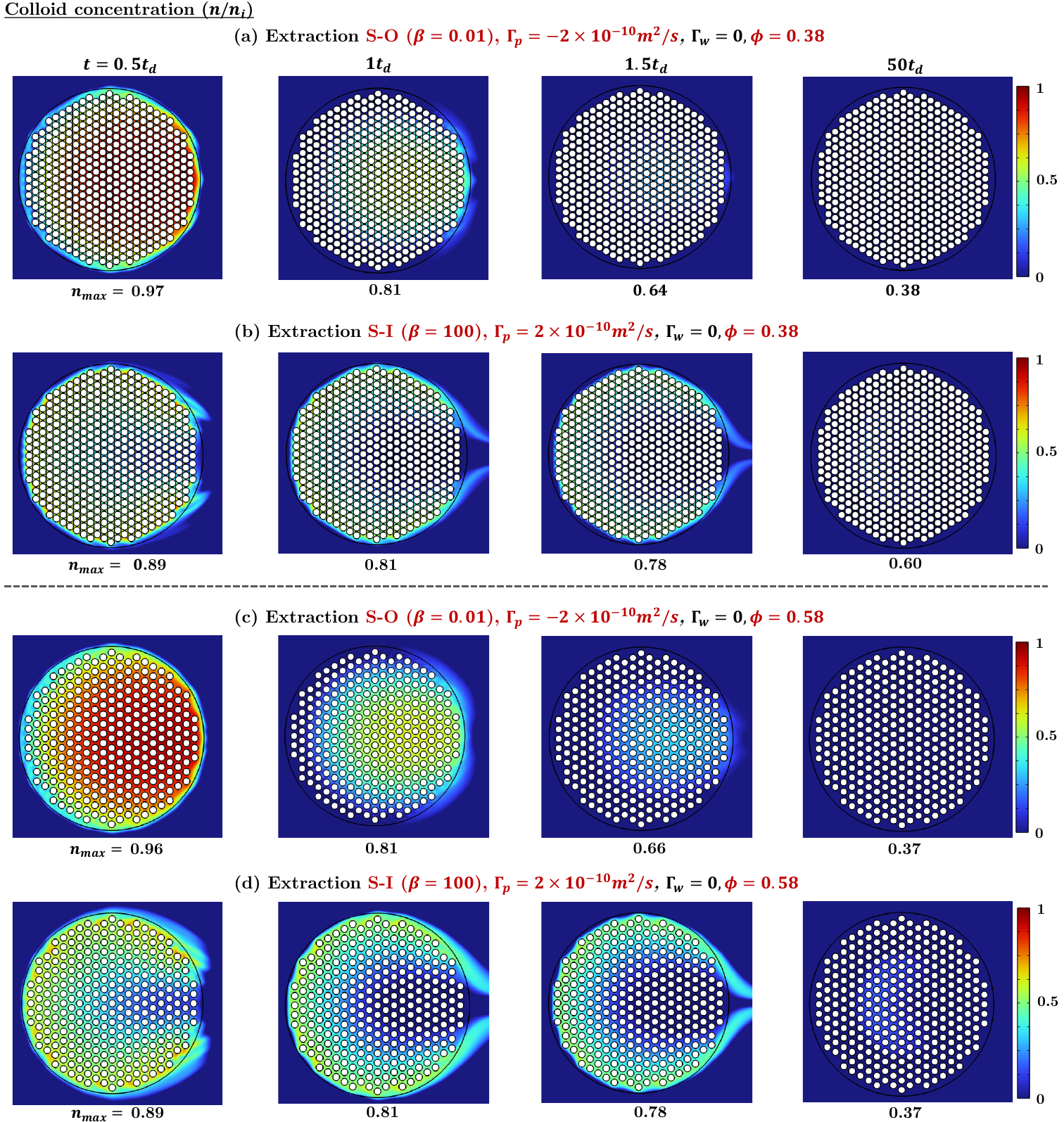}
    \caption{\small \textbf{Influence of porosity ($\phi$) on colloid extraction.} Colloid concentration variation with time for S-O and S-I modes in porous media with porosity (a,b) $\phi=0.38$ and (c,d) $\phi=0.58$. Parameters: solute ambipolar diffusivity $\mathcal{D}_s=10^{-9}$ $m^2 /s$, colloid diffusivity $\mathcal{D}_p=10^{-12}$ $m^2 /s$, colloid mobility $\Gamma_p=\pm 2\times 10^{-10}$ $m^2/s$, yarn radius $R=400$ $\mu m$, fiber radius $R_f=15$ $\mu m$, surface-to-surface gap between fibers $S_f=4.2$ $\mu m$ ($\phi=0.38$) and $11.5$ $\mu  m$ ($\phi=0.58$), number of fibers $N_f=4.39$ ($\phi=0.38$) and $301$ ($\phi=0.58$).
 }
    \label{Fig:Porosity_variation_colloid_profiles}
\end{figure}

\noindent
Fig. \ref{Fig:Porosity_variation_colloid_profiles} shows the spatial signatures of two modes (S-O and S-I) with the variation in fibre spacing ($S_f$) from $4.2$ $\mu m$ ($\phi=0.38$) to $11.5$ $\mu m$ ($\phi=0.58$), does not qualitatively alter the results and conclusions of the current work. Beyond $\phi=0.6$, the advection starts to impact colloid profiles. 

\subsection*{6. Sensitivity analysis of osmotic and phoretic mobility ratio ($\Lambda$)} 
\begin{figure}[H]
    \centering
\includegraphics[width=0.5 \textwidth]{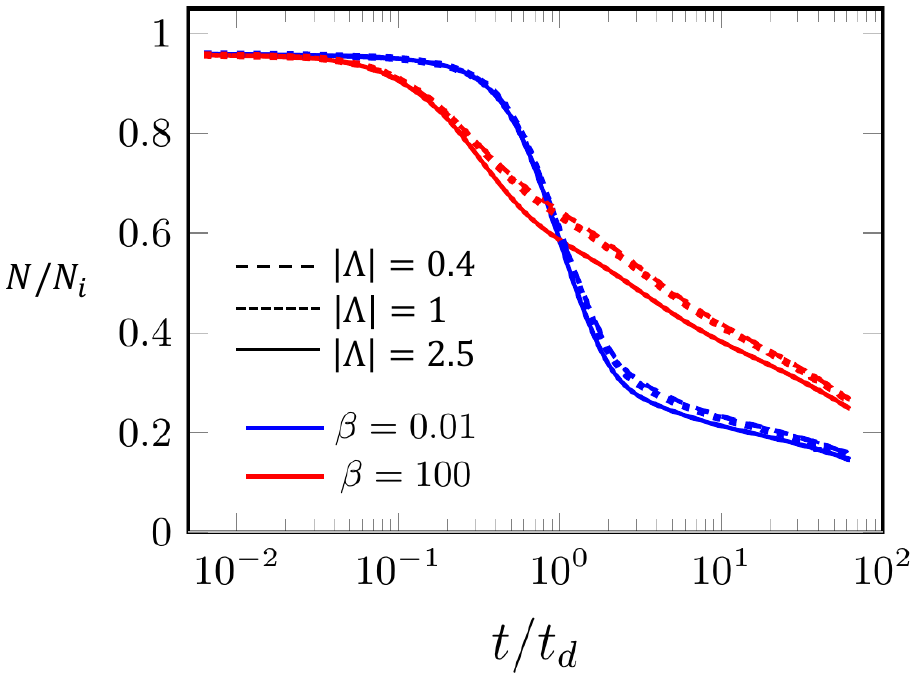}
    \caption{\small \textbf{Influence of osmotic and phoretic mobility ratio on colloid extraction.} Variation of total number of colloids (per unit length) with time for osmotic and phoretic mobility ratio $|\Lambda|=0.4$ (dashed lines), $1$ (dotted lines), and $2.5$ (solid lines) at $\Gamma_w<0$. Parameters: $\mathcal{D}_s=10^{-9}$ $m^2 /s$, $\mathcal{D}_p=10^{-12}$ $m^2 /s$, $\Gamma_p=\pm 2\times 10^{-10}$ $m^2/s$, $\Gamma_w=-5\times 10^{-10}$ $m^2/s$, $R=400$ $\mu m$ (domain radius), $R_f=15$ $\mu m$ (fiber radius), $S_f=7.5$ $\mu m$ (spacing), and $\phi=0.48$ (porosity).
 }
    \label{Fig:Gp_and_Gw_ratio3}
\end{figure}

\noindent
Fig. \ref{Fig:Gp_and_Gw_ratio3} shows the sensitivity analysis for three different values of $|\Lambda|$. 
As we increase the $|\Lambda|$, the increased internal advection (due to increased osmotic flow) modestly enhances the colloid extraction. We have additionally examined the opposite sign of wall mobility (\(\Gamma_w>0\)) in Fig. S2, which shows qualitatively similar behavior; hence we focus on one sign of the mobility here.

\end{document}